Title: Vibrating wires for beam diagnostics

Corresponding Author: Dr Suren Arutunian,

Corresponding Author's Institution: Yerevan Physics Institute

First Author: Suren Arutunian

Order of Authors: Suren Arutunian; Manvel Mailian; Kay Wittenburg, Dr

Abstract: A new approach to the technique of scanning by wires is developed. Novelty of the method is that the wire heating quantity is used as a source of information about the number of interacting particles. To increase the accuracy and sensitivity of measurements the wire heating measurement is regenerated as a change of wire natural oscillations frequency. By the rigid fixing of the wire ends on the base an unprecedented sensitivity of the frequency to the temperature and to the corresponding flux of colliding particles. The range of used frequencies (tens of kHz) and speed of processes of heat transfer limit the speed characteristics of proposed scanning method, however, the high sensitivity make it a perspective one for investigation of beam halo and weak beam scanning. Traditional beam profile monitors generally focus on the beam core and loose sensitivity in the halo region where a large dynamic range of detection is necessary. The scanning by a vibrating wire can be also successfully used in profiling and detecting of neutron and photon beams.

# Vibrating wires for beam diagnostics

*S.G. Arutunian*[1]*, M.R.Mailian, Yerevan Physics Institute*
*Kay Wittenburg, DESY*

***Abstract***

A new approach to the technique of scanning by wires is developed. Novelty of the method is that the wire heating quantity is used as a source of information about the number of interacting particles. To increase the accuracy and sensitivity of measurements the wire heating measurement is regenerated as a change of wire natural oscillations frequency. By the rigid fixing of the wire ends on the base an unprecedented sensitivity of the frequency to the temperature and to the corresponding flux of colliding particles. The range of used frequencies (tens of kHz) and speed of processes of heat transfer limit the speed characteristics of proposed scanning method, however, the high sensitivity make it a perspective one for investigation of beam halo and weak beam scanning. Traditional beam profile monitors generally focus on the beam core and loose sensitivity in the halo region where a large dynamic range of detection is necessary. The scanning by a vibrating wire can be also successfully used in profiling and detecting of neutron and photon beams.

[1] Contact person: Suren Arutunain, Alikhanian Br. Str. 2, 375036, Yerevan Physics Institute, Yerevan, Armenia, Tel.: (3741)350114; e-mail: femto@yerphi.am

### *Introduction*

One of the wide-spread methods of diagnostics in accelerators is the scanning of charged particles beams by thin wires (see e.g. [1-6]). The principle of operation is based on measurement of fluxes of secondary particles scattered off the wire, the intensity of which is proportional to the quantity of primary particles colliding with the wire. The methods to measure the fluxes of secondary radiation are fast therefore the speed of measurement is practically limited by the speed of wire motion.

In this paper a new approach to the technique of scanning by wires is developed. Novelty of the method is that the wire heating quantity is used as a source of information about the number of interacting particles. To increase the accuracy and sensitivity of measurements the wire heating measurement is regenerated as a change of wire natural oscillations frequency. By the rigid fixing of the wire ends on the base an unprecedented sensitivity of the frequency to the temperature and to the corresponding flux of colliding particles. The idea to use the vibrating wires frequency for beam diagnostics was suggested in [7, 8].

The range of used frequencies (tens of kHz) and speed of processes of heat transfer limit the speed characteristics of proposed scanning method, however, the high sensitivity make it a perspective one for investigation of beam halo and weak beam scanning [9]. Traditional beam profile monitors generally focus on the beam core and loose sensitivity in the halo region where a large dynamic range of detection is necessary.

The scanning by a vibrating wire can be also successfully used in profiling and detecting of neutron and photon beams.

### *Stretched wire oscillations frequency taking into account its elasticity*

Transverse oscillations are the main type of oscillations of a wire stretched between two clamps. In the following, oscillations in one transverse direction were assumed.

Equation of transverse oscillations of a stretched wire taking into account its elasticity is written as [10]

$$I_M E X'''' - F X'' + \rho S \ddot{X} = 0, \tag{1}$$

where $X$ is the transverse deviation of the wire, prime means derivative along the $z$ axis of the wire, point means time derivative, $I_M$ is the moment of inertia of the wire transverse cross-section (for a round wire of diameter $d$: $I_M = \pi d^4 / 64$), $E$ is the wire modulus of elasticity, $T$ is the wire tension, $S$ is the wire cross-section, $\rho$ is the density of wire material. The wire with fixed ends can be considered as a stretched vibrating wire, when the first member in (1) is small. On the contrary, when the second member is small, this object is a beam with fixed ends.

Solving equation (1) in the general case: $l$ is the wire length, $x = z/l$ is a dimensionless coordinate along the wire. Then the equation (1) can be rewritten as

$$\varepsilon \omega_0^2 X'''' - \omega_0^2 X'' + \ddot{X} = 0. \tag{2}$$

Here $\varepsilon = \dfrac{d^2}{16 l^2} \dfrac{E}{\sigma}$, $\omega_0^2 = \dfrac{\sigma}{l^2 \rho}$, $\sigma = \dfrac{F}{S}$ is the wire tension, and the prime is the derivative with respect to the dimensionless coordinate $x$. One can solve the equation (2) by the method of partition:

$$X = X_1(x) X_2(t). \tag{3}$$

The equation (2) can be written as equality

$$\frac{\varepsilon\omega_0^2 X_1'''' - \omega_0^2 X_1''}{X_1} = -\frac{\ddot{X}_2}{X_2}. \qquad (4)$$

Since the left side of this equation is function of $x$ only, and the right side of $t$ only, the equality can be achieved in case when both sides are equal to a constant. Denoting it as $-\lambda_0^2$, we obtain two equations for $X_1$ and $X_2$:

$$\varepsilon X'''' - X_1'' - \Lambda^2 X_1 = 0,$$
$$\ddot{X}_2 + \lambda_0^2 X_2 = 0, \qquad (5)$$

where $\Lambda^2 = \lambda_0^2 / \omega_0^2$.
The general solution of the first of equations (5) is written in the form

$$X_1 = c_1 \sin k_1 x + c_2 \cos k_1 x + c_3 \sinh k_3 x + c_4 \cosh k_3 x, \qquad (6)$$

where

$$k_1 = \sqrt{(\sqrt{1+4\varepsilon\Lambda^2} - 1)/2\varepsilon},$$
$$k_3 = \sqrt{(\sqrt{1+4\varepsilon\Lambda^2} + 1)/2\varepsilon}.$$

Put natural boundary conditions on solution of (6):

$$X_1(0) = X_1'(0) = X_1(1) = X_1'(1). \qquad (7)$$

Nonzero constants in expression (6) are achieved in case when the constant $\lambda_0$ is the solution of characteristic equation:

$$2\frac{k_1}{k_3}\frac{1}{\cosh k_3} - 2\frac{k_1}{k_3}\cos k_1 + (1 - \frac{k_1^2}{k_3^2})\sin k_1 \tanh k_3 = 0. \qquad (8)$$

This degeneracy condition defines the dependence of the oscillations frequency on the task parameters. In case of $\varepsilon << 1$ the studied object is a stretched vibrating wire. In this case is $k_1 \approx \Lambda, k_3 \approx 1/\sqrt{\varepsilon} >> 1$ and we can neglect the first member in (8), and put $\tanh k_3$ and obtain:

$$(1 - \frac{k_1^2}{k_3^2})\sin k_1 - 2\cos k_1 = 0. \qquad (9)$$

The first harmonics value by decomposition of $\Lambda$ near $\pi$ can be obtained by:

$$\Lambda \approx \pi(1 + 2\sqrt{\varepsilon}), \qquad (10)$$

and the corresponding frequency of the first harmonics of the wire oscillations can be written in the form:

$$f_1 = \frac{1}{2l}\sqrt{\sigma/\rho}(1+\frac{d}{2l}\sqrt{E/\sigma}). \quad (11)$$

One has to take into account the wire material's elasticity results in increment of wire oscillations frequency by the value $(d/4l^2)\sqrt{E/\rho}$, which is independent of wire tension.
Similar calculations can be done for non-round wire, having two different moment of inertia of transverse cross-section $I_a$ and $I_b$. Corresponding frequencies along the main axes of inertia are written as

$$f_1^a = \frac{1}{2l}\sqrt{\sigma/\rho}(1+\frac{\sqrt{2}I_a^{1/4}}{\pi^{1/4}l}\sqrt{E/\sigma})$$
$$f_2^b = \frac{1}{2l}\sqrt{\sigma/\rho}(1+\frac{\sqrt{2}I_b^{1/4}}{\pi^{1/4}l}\sqrt{E/\sigma}) \quad (12)$$

For a rectangular wire of cross-section $d_a \times d_b$ $(d_a$ $(d_a \approx d_b)$, the difference of frequencies of transverse oscillations ($I_a = d_a d_b^3/12, I_b = d_b d_a^3/12$)

$$\Delta f = \sqrt[4]{1/16\pi}\sqrt[4]{\Delta d/d}(d/l^2)\sqrt{E/2\rho}, \quad (13)$$

where $d \approx d_a \approx d_b$, $\Delta d = d_a - d_b$.

For example, for a wire of length about 30 mm and diameter about 50 μm, the non-roundness of the wire 0.1 μm results in frequency shift of different transverse modes about 4 Hz. Usually the wire oscillations are excited by interaction of the current trough the wire and magnetic field. Normally this system aimed to generate only one transverse mode. But in practice because of some inaccuracy of magnetic fields there is some coupling between these transverse modes. Increment of the difference between frequencies of two transverse modes decreases the possibility of capture of one mode oscillations by the second one. The difference can be increased e.g. by flattening of the wire. In this case the difference become few hundreds Hz and capture of oscillation in second mode does not occur. However, it is a constant shift in the frequency and will not contribute to the relative measurements of beam tails or profiles. Moreover a big difference between the modes is preferable for separation of transverse modes.

***Wires oscillations excitation***

Exciting action on the wire arises as a result of interaction between the current through the wire and magnetic field of the permanent magnets. When the current passes through the wire a force of interaction shifts the wire in transversal direction by the value proportional to the magnitude of the magnetic field. In its turn, time-dependent wire shift induces an electromagnetic force on wire ends. The value of this voltage is proportional to the time derivative of the current, which from the electrical viewpoint is equivalent to an inductance. So the electrical equivalent of the wire can be described as a series of a resistor and inductance (see e.g. [11]).

Thus, excitation of mechanical oscillations in wire is possible because the wire acts like a tuned circuit when placed in an amplifier feedback arrangement.

When the wire is connected in an amplifier circuit, Fig. 1, a small amount of energy is fed back to the wire, which causes it to vibrate. This is similar to the excitation of electromechanical oscillations in quartz resonators (see, e.g. [12])

*Fig. 1. Schematic layout of wire oscillations generation with usage of amplifier with feedback.*

The wire is connected in positive feedback circuit (R1) and is the lower arm of the divider; the output of the divider is connected to the noninverting input of the operating amplifier. This connection supports the frequency of oscillations.

To support the amplitude of oscillations in some definite level an amplitude comparator on the base of operational amplifier is provided and connected to the negative feedback circuit (R2). As a control voltage serves the amplified difference between the amplitude of negative half-period of oscillations in generator's output and reference voltage of the potentiometer. Amplification factor of this difference is of the order $1 \div 5\times10^5$, and hence the amplitude is maintained practically with absolute accuracy.

***Temperature dependence of oscillations frequency***

Oscillation frequencies of a vibrating wire based electromechanical resonator have strong temperature dependence. Roughly the electromechanical resonator can be represented as a base made of a material with a coefficient of thermal expansion $\alpha_B$ and with a wire with corresponding factor $\alpha_S$. The rigidity of the base is much more than that of the wire hence the wire length is defined by the distance between clips of the base. Suppose the electromechanical resonator was assembled at a certain temperature $T_0$, when the length of unstrained wire was $l_{S0}$ and the distance between points of wire ends fixation was $l_{B0}$. Thus at the moment of the assembling initial strain of the wire was $\sigma_0 = E(l_{B0} - l_{S0})/l_{S0}$. In the general case of electromechanical resonator operation we can suppose that the wire temperature has an average value $T_S$ along the wire, and the average temperature of the base is $T_B$. The relative change of the wire strain is defined by the expression:

$$\frac{\sigma}{\sigma_0} = \frac{E}{\sigma_0}\frac{(l_{B0} + \alpha_B(T_B - T_0)) - (l_{S0} + \alpha_S(T_S - T_0))}{l_{S0}} = 1 + \frac{E}{\sigma_0}\frac{\alpha_B(T_B - T_0) - \alpha_S(T_S - T_0)}{l_{S0}}. \tag{14}$$

The relative change of the wire geometrical sizes are:

$$\frac{l_S}{l_{S0}} = 1 + \frac{\alpha_S(T_S - T_0)}{l_{S0}}. \tag{15}$$

A comparison of the formula above with (14) shows that the main reason of the electromechanical resonator frequency shift is the change of wire strain, because the dimensionless quantity $E/\sigma_0$ is great even in case when $\sigma_0$ is close to the maximal allowable strain of the wire $\sigma_P$. In Table 1 some characteristic values of $E/\sigma_{0.2}$ ($\sigma_{0.2}$, is a tension, for which the residual deformation after removal of the load is 0.2 % ) for some materials are presented [13, 14]. In Table 1 we present also the temperature range of the vibrating wire sensor for tension $\sigma_{0.2}$:

$$\Delta T_S = \sigma_{0.2}/\alpha_S E \tag{16}$$

*Table 1. Temperature range of vibrating wire sensor.*

The value of $\alpha_s$ is taken at a temperature of 600 K. In data marked by * sign the half of the breaking point is taken as $\sigma_{0.2}$. From the table is seen that Titan alloys and SiC threads hold much promise. However, in the latter case the problem of dielectric wire oscillations excitation is to be solved.

*Fig. 2. Dependence of the vibrating wire frequency on the wire temperature for different tensions of the wire.*

Fig. 2 helps to choose the geometric parameters of the wire and its strains for the electromechanical resonator making. Here the dependence of the beryl bronze wire oscillations frequency (first harmonics, correction of elasticity member is neglected) on temperature are presented for different values of wire strain at 300 K. The strain range is chosen from 100 Mpa (lower line in Fig. 2) to 350 Mpa (upper line). The wire breaking point is about 450 Mpa. For strains more than 350 Mpa some problems with stability of the wire mechanical parameters arise and for strains less than 100 Mpa difficulties with excitation of the wire oscillations arise. At heating the wire strain decreases and the wire oscillations frequency also decreases. It is difficult to excite oscillation below 1 kHz, therefore the frequency range is cut by this value. As can be seen form the Fig. 2 the maximal temperature range for beryl bronze wire operation is about 120 K. However, such a resonator has a very high sensitivity to the temperature. By using tungsten wires it is possible to increase the temperature range as a consequence of less of thermal expansion coefficient of tungsten, as well as because the limit strain of tungsten three times less than that of beryl bronze. Thus choice of wire material allows optimizing the sensitivity and operating range of pickup.

### *Wire under beam irradiation*

This chapter discusses the situation when the wire is irradiated by a high energy particles or photons beam. The interaction of the beam with the wire mainly causes heating of the wire due to the energy loss of the particles in the wire. Therefore the frequency of natural oscillations of the wire will provide information about its temperature. For charged particles the main parameter of energy transfer is the coefficient of ionization loss dE/dy, which is described by the Bethe-Bloch-formula. A major part of the energy is spent on processes of emitting secondary particles. Measurements at CERN and DESY [2] had shown that only about 30% of the energy is transferred into.

### *Some simple estimates*

The average ionization loss of a particle passing though the wire $q_1$ (eV) is:

$$q_1 = k\frac{dE}{dy}\frac{\pi}{2}r \qquad (17)$$

The heating depends on transformation ratio of ionization loss into heat $k$ (we suppose $k = 0.3$), ionization loss of a particle is $dE/dy$ and wire radius is $r$ ($d' = \frac{\pi}{2}r$ 'mean radius').

The current of particles deposited on the wire $I_S$ (A) is defined by the expression:

$$I_S = I_0 \cdot \int_{x-\frac{d'}{2}}^{x+\frac{d'}{2}} \frac{1}{\sigma_x \cdot \sqrt{2\cdot\pi}} \cdot \exp\left(-\frac{x^2}{2\cdot\sigma_x^2}\right)dx \qquad (18)$$

where $x$ is the distance between the wire and the beam center. The beam has a Gaussian distribution with horizontal size $\sigma_x$ and $I_0$ is the beam current.

The power $Q_S$ heating the wire is: $Q_S = q_1 I_S$. Due to thermal losses on thermoconductivity (the main cooling mechanism at room temperature) $Q_S$ causes the wire heating with respect to environment temperature by the value

$$T_{mean} = g_F \frac{Q_S l}{8\pi r^2 \lambda}, \tag{19}$$

where $\lambda$ is the coefficient of thermal conductivity of the wire, $q_1 = k(dE/dy)(\pi/2)$, $k$ is transformation ratio of ionization losses into heat, $dE/dy$ is the particle ionization losses, $r$ is wire radius, $I_S$ is current of particles in vertical gap between magnet poles heating the wire, form-factor $g_F \approx 1$ determines the difference between the triangle model of temperature profile and exact solution (see below).

$$\Delta f = -0.25 f(E/\sigma)\alpha_S T_{mean}. \tag{20}$$

Some typical parameters (experiments at PETRA, detailed see below): beryl-bronze wire with $l$ = 36 mm and $r$ = 0.045 mm, $\lambda$ = 0.17 W/K/mm and $\alpha_S$ = 1.9 $10^{-5}$ $K^{-1}$. Parameter $q_1$ is equal to 27.57 keV ($dE/dy$ = 1.3 MeV/mm in bronze). In case of $f$ = 4900 Hz that corresponds to the wire strain $2.801 \times 10^8$ Pa and the ratio $E/\sigma$ is equal to 464. Table 2 presents the results of the estimation of $I_S$, $Q_S$, $T_{mean}$ and frequency decrement $\Delta f$ at different positions of the wire with respect to beam center for protons of energy 15 GeV, current $I_0$ = 10 mA and $\sigma_x$ = 6.083 mm.

*Table 2. Some typical parameters of wire heating process by proton beam.*

The area > $2.5\sigma$ of a beam is often called beam halo or "tails" [15]. As is seen from the Table 2 this region is covered by the VWS sensitivity range.

***The characteristics of the wire heating process***

In the ultra high vacuum of an accelerator chamber the wire's thermal balance is determined by the following processes: heat deposit in the wire material caused by particle dispersion, heat transfer caused by thermal conductivity of the wire's material and its fixing system, radiation of heat from the wire's surface. When the particles flux is constant (fixed position of the wire or its slow motion) the average equilibrium temperature can be determined according to balance of the coming power and cooling effects. Assuming a relative low temperature increase of the wire in a beam halo, the main cooling effect is the heat transport along the wire. (The black body radiation (~$T^4$) is in the considered range of parameters nearly negligible).

The considered object is a stretched wire therefore the temperature dependence on the cross coordinates of the wire can be neglected. Transition to the one-dimensional model is done in following way. On heat sources: The whole power deposited on the wire as result of wire and beam interaction and the space density of the sources inside the wire were calculated. We consider that the space density is uniform in the cross-section of the wire. The same is done for the black body radiation. In a uniform temperature distribution model along the cross-section of the wire we calculate radiation losses taking into account the real edges of the wire and normalize it with respect to the wire volume. The mentioned expressions are then inserted into the general 3-dimensional Laplace equation and the dependence on the cross coordinates of the wire are neglected. Choosing the following co-ordinates: $z$ axis – along the wire direction, it is also the vertical cross coordinate of the beam, $x$ axis – horizontal transverse coordinate, $y$ axis is the beam spread direction. In this case the heat conductivity equation taking into account heat dispersion will be written in a following form:

$$c\rho\frac{\partial T}{\partial t}=\frac{\partial}{\partial z}\left(\lambda\frac{\partial T}{\partial z}\right)+q-\frac{2k}{r_w}(T^4-T_0^4). \tag{21}$$

Here $c$ is the coefficient of heat conductivity, $\rho$ is the density and $\lambda$ is the coefficient of heat conductivity of the wire's material, density of the heat source, $k$ is the Stefan-Boltzman constant, $T=T(t,z)$ is the temperature of the wire along the $z$ axis depending on time, $q$ is the density of the heat source, $r_w$ is the radius of the wire,.

The transverse density of the current is modeled by Gaussian distribution:

$$I=t_f\cdot I_0\frac{1}{2\pi\sigma_x\sigma_z}exp\left(-\frac{x^2}{2\sigma_x^2}\right)exp\left(-\frac{z^2}{2\sigma_z^2}\right) \tag{22}$$

The bunch structure of the beam is modeled by the depending on time $t_f$ multiplier, which is equal to 1 in the limits of the bunch and equal to 0 in interbunch space, $I_0$ is the average current in the bunch, $\sigma_x$ is the width of a bunch along the $x$ direction and $\sigma_z$ - along the $z$ axis. The calculations experience showed that in many cases the bunch structure of the beam could be neglected. In that case it is necessary to set $t_f=1$, and $I_0$ will be the current mean value, averaged by bunches.

In general for the beam, divided into bunches, the value of $q(x,z)$ we calculate in following way

$$q(x,z)=t_f\cdot z_f\cdot K_{tr}\cdot\frac{dE}{dy}\cdot I_0\cdot\frac{2r_wh_z}{2\pi\sigma_x\sigma_z}exp\left(-\frac{x^2}{2\sigma_x^2}\right)exp\left(-\frac{kh_z}{2\sigma_x^2}\right)\cdot\frac{1}{\pi r_w2h_z}. \tag{23}$$

We consider that in x direction in the range of the cross-section of the wire the density of the current change is small and the function (22) is substituted by the value of the current in the middle of the wire. In the formula (23) $z_f$ is the factor, which takes into account the screening of the vibrating wire by magnetic system and clamps, $dE/dy$ is the value of the ionization losses of the charged particles on a wire's material, $K_{tr}$ is the factor of transformation of ionization losses into heat, $h_z$ is the step of the numerical scheme along the $z$ axis.

As we are going to solve the equation (21) numerically, we are also considering temperature dependencies of the equation factors – heat capacity and heat conductivity of the wire.

A method analogous to Krank-Nickolson method [16] to solve the equation of heat conductivity with constant factors was used to solve equation (21). The dependence of the thermal conductivity and thermal capacity factors from temperature are modeled by polylines.

A numerical program including a graphical interface was written to calculate the given equations. Fig. 3 presents the temperature profiles along the wire for different times after a wire irradiation by a 15 Gev proton beam with a mean beam current 1 mA. It was taken into account that some space of the wire is covered by magnets system due to some real mechanical limitations. The beam width $\sigma$ in horizontal direction was set to 6 mm. One can see that the temperature distribution trends to a triangle.

*Fig. 3. Irradiated wire temperature profile calculations at different times: 1- after 1 sec, 2 – after 2 sec etc till 7 sec and then for 10 sec. By bold curve presented power of radiation by 1 mA proton beam ($\sigma_x$ = 6 mm).*

Parameters of the wire: material – Beryl-Bronze, diameter – 90 µm, length – 36 mm. Wire placement – 20 mm from beam center in vertical direction.

*Fig. 4. Dynamics of temperature of the wires central point – 1; wire natural frequency (second harmonic) – 2. The initial frequency of the wire is 4850 Hz.*

The dynamics of the wire maximal temperature and corresponding shifts of frequency are presented in Fig. 4. One can observe that the maximal shift of frequency trends to approximately 40 Hz.

The temperature profiles of wires of different materials are calculated to compare the behavior of different materials. The geometrical parameters were always the same: The diameter is 90 mm, length is 36 mm. In these calculations the wire heating process is modeled by a 15 GeV 1 mA proton beam with $\sigma_x = 6$mm. The wire is placed at $3.3\sigma_x$ from the beam center. Fig. 5 presents the results of such calculations for Titan, Beryl-Bronze, Platinum and Tungsten.

*Fig. 5. Temperature profile in the wire for different materials.*

As seen from the Fig. 5 the maximum temperature shift is achieved for a Pt wire. However due to the larger thermal expansion coefficient the frequency shift is largest for a Ti wire (see Table 3).
In Table 3 we also present the energy losses through the radiation mechanism and thermal conductivity.

*Table 3. Charged particles energy losses through the radiation mechanism and thermal conductivity.*

Here Qrad are the energy losses by radiation (last term in equation (21)), Ql are the energy losses by thermal conductivity mechanism, T(0) – is the temperature increasing at central point of the wire and DF is the corresponding shift of frequency in case when secon harmonic of oscillations is generated. One can see that for present parameters range Qrad<<Ql.

The frequency shifts for wires from different materials are in the same order so more important are the stability of materials mechanical parameters

***Photon beams***

For photons and neutral particles the beam heating process is described by similar physical parameters.

Photon beam passing through the material will also cause heating of the wire material. Some estimates of absorption parameters $\mu^{-1}$ (cm$^{-1}$) for different materials and photon energies are presented in Table 4. Few words about calculation method with Ref. – will be done!

*Table 4. Absorption parameter $\mu^{-1}$ (cm$^{-1}$) for photon beams with different energies.*

To estimate the wire frequency shift under irradiation of photon beam we numerically solved the model task of thermal conductivity along the wire allowing finding the temperature profile along the wire for different photon beam parameters, wire materials and geometry.

For example we consider the photon beam with parameters of XFEL TESLA [17]:

Wavelength, 1-5 A

Average power, 210 W

Photon beam size (FWHM), 500 μm at distance 250 m from source

Average flux of photons $1.0\times10^{+17}$ ph/s.

Vibrating wire parameters:

Length about 10 mm,

Diameter 20 μm,

Initial frequency (the second harmonics) - 5000 Hz.

Fig. 6 represents the temperature profiles of wires from Molybdenum and Silicon at different time moments [18]. It is seen that the temperature balance is about 1 sec. Note that the temperature profile has almost a triangle shape because the beam size is much less of wire length.

*Fig. 6. Dynamics and profile of temperature along the wire from Silicon and Molybdenum. The parameters of the photon beam are the same and the VWS is placed at distance* $5\sigma$ *from the beam center in both cases.*

As seen from Fig. 6 the sensitivity of the VWS based on the Molybdenum is about -58 Hz/K and -18.6 Hz/K for Silicium. It means that for this range of photon energy the Molybdenum wires is preferable.

The lower limit of the sensitivity of the VWS to temperature changes is defined by electronic noise and the method of signal measurement and is less than 0.01 Hz, which corresponds to temperature resolution of $2\times10^{-4}$ K for VWS based on Molybdenum wire. The temperature determines the upper limit, when the wire oscillation generation broke out. This corresponds to approximately 3000 Hz, while the temperature shift is about 54 °C. This value is determined by the preliminary tension of the Molybdenum wire.

At a power of less than $1\times10^{-9}$ W the wire frequency shift becomes unresolvable (the corresponding distance to the beam center is $6.4\sigma$), at a power of about 0.1 mW (the distance is $4.2\sigma$) the frequency shift is about 330 Hz (the second harmonics) and corresponding heating of the wire is about 6 $^0$C.

### *Vibrating Wire scanners, Experimental results*

In the vibrating wire sensor the generation of wire's vibrations is achieved by means of interaction of the alternating current of about 1 mA passing through the wire with the permanent magnetic field. Strong samarium-cobalt magnets are used, which provide a field strength of order of > 8 kGs in the working range. If the field is smaller, the generation of oscillations becomes unstable.

A magnetic field scheme is used which possesses a minimum interference between beam and magnetic field of scanner. The central part with sizes of about 14 mm is left free for beam scanning. The main view of VWS developed for PETRA is presented in Fig. 7.

*Fig 7. VWS main view with part description: 1 - one of the clips (is fastened on the base butt-end), 2 - the sensor base made of material with low coefficient of thermal expansion, 3, 5 – magnet poles system, 4 – dielectric plates, 6 - vibrating wire, 7, 8 – fasteners.*

TheVWS were used also for laser beam and for ion beam profiling with some small modifications [19, 20].

***Photon beams, Laser beams***

Scanning of continuous and pulse He-Ne, YAG:Er, YAG:Nd laser beams yields following results. The resolution with respect to radiation power density was achieved of order $5\times10^{-4}$ W/cm$^2$ (632.8 nm, spatial resolution ~40 µm, see Fig. 8). Power measurement threshold was ~$10^{-5}$ W, that of energy (for pulse lasers) ~$10^{-5}$ J, at the linear dynamic interval of radiation intensity of the pickup greater than 103. Time of achieving of frequency to the 95 % of maximal value was ~1 sec.

*Fig. 8. Laser beam scanning by vibrating wire.*

The wire motion was made step-by-step manually.

An interesting result was obtained when the dose of irradiation by photon beam is small [18]. As a result of such irradiation the wire material can undergo some long-term-changes and hence the parameters of the frequency drift and the dependence of frequency on temperature had also change on a long-term-base. Fig. 9 represents the characteristics of the frequency change versus temperature for two wires, one (the lower curve) was 45 minutes irradiated under a photon beam of energy of 100 keV of X-rays with an intensity of about 2000 Roentgen/min.

*Fig. 9. Frequency-temperature dependence of wires before (a) and after (b) irradiation.*

The X-ray radiation resulted in a change of the temperature dependence of the irradiated wire from 7.03 Hz/K (before irradiation) to 4.7 Hz/K (after), while for the non-irradiated wire this value changed negligibly (3.26 and 3.1 Hz/K, correspondingly). Thus, already small doses of radiation can influence the mechanical tensions and/or redistribution of dislocations, and this was fixed by our pickup. Method of irradiation of the wire by X-rays can also be used for studies of radiation quenching and aging of materials. Note, that investigations in this area by traditional methods require long-term experiments, while high sensitivity of the vibrating wire pickup allows to fix the changes practically on-line. So vibrating wire electromechanical resonators can be used for express analysis of characteristics of materials, irradiated by photon and particles beams in wide range of parameters.

***Electron beam***

The first scanning experiments on a charged beam were done on an electron beam at the Injector of Yerevan Synchrotron with an average current of about 10 nA (after collimation) and an electron energy of 50 MeV.

Fig. 10 represents the result of the reconstruction of the beam profile for the first half-distance scanning. The solid line represents the profile of the beam approximated by the mean square method of a Gaussian function with a standard deviation of $\sigma$ = 1.48 mm and a beam position at 30.87 mm. The overall current of the beam was measured to $I_0$ = 10 nA. A more detailed description of the experimental results can also be find in [21-24].

*Fig. 10 Horizontal profile of an electron beam with a current of 10 nA scanned with a beryl-bronze wire of 90 µm diameter as the vibrating wire sensor.*

***Ion beam***

The vibrating wire scanner (VWS) was also tested on an ion beam of energomassanalyzer EMAL-2 [18, 24]. The Ion beam current of about 1 nA and an energy of 20 keV was obtained. A wire of diameter 90 μm was used and the frequency was measured at the central position of the beam (see Fig. 11).

*Fig. 11. VWS under ion beam of EMAL-2.*

The ion beam was locked by a diaphragm and only in a period between 15 and 30 seconds the ion beam hit the wire. A part of about 16 pA beam current interacted with the wire and a frequency decrement of about 0.15 Hz was measured.

***Proton beam***

A series of experiments with the VWS were done on proton beam of the accelerator PETRA at DESY. The place for VWS location was chosen in the so-called "proton by-pass" to avoid any influences or destructions from higher order mode losses created by short electron bunches which are accelerated in the same machine. The beam was specially prepared for our purposes and consisted of 10 bunches with initial current about 15 mA and an energy 15 GeV. A system of two scintillator-photomultiplier pickups (PM1 and PM2) were installed additionally to measure particles scattered on the wire.

The main view of the VWS in its parking position is presented in Fig. 12.

*Fig. 12. VWS in park position in PETRA vacuum chamber. Wire distance from beam center is* $6.7\sigma$ *(40 mm).*

The origin of the coordinate system is placed in vacuum chamber center and the horizontal axis is directed to the right to the center of the accelerator.

Color gradation corresponds to the fall of beam flux density by one order (in $s^{-1}cm^{-2}$ units). Beam parameters are following: I = 10 mA, $\sigma_x = 0.6$ cm, $\sigma_z = 0.5$ cm. The ellipse in the center corresponds to the beam particles density $10^{16}$ - $10^{17} s^{-1}cm^{-2}$. It is seen that in this position particles flux density falling on the wire is by two orders less than flux falling on the corner of magnet pole (see also Fig. 7 with main view of the VWS).

The parking position of the wire is located at the 40 mm from the axis of the vacuum chamber. The movement of the scanner is done by a stepping motor. Also a system of adjustable beam bumps allows to regulate the position of the beam inside the vacuum chamber.

The 2/3 part of the operating oscillating wire is covered by the magnetic system of the scanner. This scanning system is based on strong permanent magnets and specially designed to diminish magnetic field in the region of the beam. Actually the magnetic poles come out from the line of vibrating wire fairly and great number of particles can scatter on it.

Even in park position the VWS actively reacted to the events in vacuum chamber: slow changes of environment temperature, proton beam turn on/off and change of beam parameters. Fig. 13 shows a typical picture at currents less than 50 mA (at beam currents exceeding 100 mA the frequency signal became noisy).

*Fig. 13. The VWS output signal in park position.*

In this experiment the mean current of the beam was less than 50 mA. One can see a strong correlation between the beam current and the change of the frequency, even in the parking position of the scanner. The change of the frequency was in range of 70 Hz. The decrement of the frequency is about 1 Hz/mA.

The frequency measurement was done by such a method. Some quantity of periods of measured oscillations was filled by short impulses from quartz generator and by these values the period of wire oscillations was found. This method allows to calculate frequency of the wire oscillations even during the one period of wire oscillations.

In Fig. 14 a beam scanning of a displacement of the wire up to 25 mm is presented. During the beam scanning vertical correction of the beam was done. The position of the beam according to horizontal beam position monitor SL1MO was -7 mm. At the beginning of the experiment the beam vertical beam position according to the same monitor was +2.1 mm. At 19:43 the beam was shifted to the vertical position of +3.2 mm (the readings of the monitor SL37MO was about 0, SR37MO was about 1.4 mm) and at 19:50 the vertical beam position was again changed to 0 mm. Scanning was mainly done in range 20 - 25 mm. Approaching the 25 mm wire-position was accompanied by a decrement of the beam current. The reason of that was probably a strong beam scraping of the solid material at the ends of the wire. The gap in between of this prototype was to small for the present beam dimensions of PERTA. During the experiment the current decreased from 13.8 mA to 1.1 mA. Also displayed in Fig. 14 is the signal from two different adjacent scintillation counters ("beam loss"). Note the strong correlation of these signals with the change of the frequency. Even that most of the losses might came from the interaction of the beam with the solid wire support; it is a nice indication of the response of the frequency change due to beam – wire interactions.

*Fig. 14. Beam scanning at full feed with different vertical shifts of the beam: 1 – the frequency signal from VWS, 2 – position of the VWS, 3 – beam current, 4 – beam losses.*

At the beginning of the experiment, simultaneously with the movement of the scanner to the position 20 mm, the frequency decreased by more than 20 Hz. An interesting effect occurs on the first scanning at position 20 mm, when the frequency after rapid falling began to rise. Probably it is connected to the fact that here the beam current began rapidly decrease due to the beam scrapping by VWS. This can be seen also on the scintillation counters. Similar behavior is observed also on other steps. Further movement of the scanner led to the essential decrement of the frequency by about 150 Hz.

Another scanning with a fixed beam position (position of the beam according to monitor SL1MO z = +3 mm, x = -7 mm) is presented in Fig. 15.

*Fig. 15. Scanning of the beam with fixed bumps. Curves: 1 – frequency signal from VWS, 2 – VWS position, 3 – beam current, 4 - beam losses.*

Beam current decreasing steps obviously are connected with scanner positions of about 15 mm. At first scanning from 10 mm to 20 mm and back at positions bigger than 17 mm frequency increased up to 5070 Hz and became noisy. This event is marked by arrows on the picture. At next scans at 20 mm at less currents of the beam such disturbances of frequency behavior did not occur.

*Fig. 16. Noises of frequency signal from VWS in stationary position.*

As it was mentioned above in stationary position as a rule the frequency smoothly drifts caused by change of beam current and environment conditions. Fig. 16 shows the deviation of the frequency from the linear approximation in position 10 mm during a 3 minute time interval (between 10:16:00-10:19:00). The Root-mean-square deviation of this sample is about $\sigma$ = 0.012 Hz. The corresponding change of the wire mean temperature is $\sigma_t$ = 0.0005 K.

Below the results of beam scanning with additional measurements of secondary particles at beam scattering on the VWS by the scintillation counter PM1 are presented in more detail.

Fig. 17 shows the results of a scanning of 12 mm displacement together with the readings of PM1.

*Fig. 17. Some shallow scans. Curves: 1 – frequency signal from VWS, 2 – VWS position, 3 – beam current, 4 – PM1 countings.*

The beam position was: z = +3.1 mm, x = -7.6 mm (readings of monitor SL1MO). One shallow scanning up to the wire position 2.5 mm and three consequent scans up to 12 mm were done. From the Fig. 17 it is seen that pickup PM1 showed signals only when approaching the position 10 mm. The entire experiment lasts about 1.5 hours and during this period the beam current monotonously decrease from 14.5 mA to 7.4 mA. No additional beam current losses were observed as a result of the larger distance between the wire and the beam center. The deep scanning to distances less than 15 mm were accompanied by correlated beam losses (see Fig. 14, 15). In accordance with overall decrease of the current a monotonous increase of the frequency of the scanner in park position as well as at 12 mm occurs.

Outcomes of VWS at positions 12 mm accompanied with spikes of frequency signal. Similar effect (more blurred) is observed in data obtained from pickup PM1 and therefore the assumption about the effects is caused by mechanical stoppage of the scanner is not valid. These spikes may be caused by result of beam halo scraping at wire supports.

An other example is shown in Fig. 18. The scan was started at a distance of 40 mm from the vacuum chamber center. Again the signal from the VWS sensor changes from the beginning of movement, while the signals from scintillators start to increase first at distances of 27 mm from the vacuum chamber center.

*Fig. 18. Scan of the proton beam using VWS: 1- frequency of the VWS, 2 – beam current, 3 – VWS position relative to the vacuum chamber center, 4 and 5 – signals from scintillators.*

The scanner was moved from park position toward the vacuum chamber center up to 20 mm. In this experiment the proton beam was shifted towards the scanner park position by distance of 4 mm by means of a local beam bump.

As seen from Figs. 17, 18, the signal from VWS appears at distances 27-40 mm from the vacuum chamber center while there is no signal from the scintillators here. Some contribution in wire heating might occure from the influence of electromagnetic higher order modes accompanying the proton beam. These electromagnetic components might are able to heat the wire by absorbing some modes (see [25]) . Clarification of this problem and corresponding modifications of VWS require additional efforts.

*Fig. 19. Comparison of signals from VWS with PM1(curve 1) and PM2 (curve 2).*

Despite from this effect, the signals from PM1 and PM2 are strongly correlate with frequency signal from VWS (see Fig. 19). Some hysteresis occurs which is connected with different direction of the VWS movement.

The largest shift of the wire oscillation frequency due to heating was measured to about 150 Hz, at a distance between the wire and beam center of x slightly less than 20 mm. This value is about a factor 2 less than the calculated value given in Table 2. Three effects might contribute to the uncertainty: 1) the uncertainty of the absolute beam position at the VWS of ±1.5 mm; 2) nongaussian beam tails; 3) electromagnetic background.

The heating of the wire by electromagnetic component of the proton beam seems to be not negligible at very far distances. We expect, that this effect will be much more dominant in the case of the much shorter bunches of an electron beam, but much less in case of DC beams. To determine this effect additional experiments are required.

### *Conclusion*

Experiments with the VWS show that the system is sensible to the beam current changes even in park position even at distance of 6.6 σ (here 40 mm) from the beam centre. The scanning shows a strong correlation between scanner position and frequency signal. The accuracy of the frequency measurements achieved 0.01 Hz which corresponds to wire temperature measurement accuracy less than 0.001 K. Comparison of the signal from VWS with a scintillating counter system registering secondary particles/radiation showed that the signal from VWS appears at least 10 mm farther from the beam centre than mentioned scintillators.

The obtained data show that the VWS can be used for beam diagnostics in accelerators. Some improvements of the scanner construction to increase the wire operating zone and to diminish the sizes of the wire support of the VWS will be necessary. At very large distances from the beam the heating of the wire due to higher order mode coupling disturbs the signal. Some more developments are necessary to get rid of this effect. Very perspectives will be the development of sensors on basis of dielectric strings, including new type of oscillations exciting and data acquisition. The area of VWS applications can be enlarged, including profiling and positioning of photon beams from synchrotron light sources and laser beams. In this case the electromagnetic coupling does not exist.

### *Acknowledgments*

The authors are grateful to PETRA stuff for friendly help during the experiments on PETRA.

## *Tables*

### *Figures*

Fig. 1. Schematic layout of wire oscillations generation with usage of amplifier with feedback.

Fig. 2. Dependence of the vibrating wire frequency on the wire temperature for different tensions of the wire.

Fig. 3. Irradiated wire temperature profile calculations at different times: 1- after 1 sec, 2 – after 2 sec etc till 7 sec and then for 10 sec. By bold curve presented power of radiation by 1 mA proton beam ( $\sigma_x$ = 6 mm).

Fig. 4. Dynamics of temperature of the wires central point – 1; wire natural frequency (second harmonic) – 2. The initial frequency of the wire is 4850 Hz.

Fig. 5. Temperature profile in the wire for different materials.

Fig. 6. Dynamics and profile of temperature along the wire from Silicon and Molybdenum. The parameters of the photon beam are the same and the VWS is placed at distance $5\sigma$ from the beam center in both cases.

Fig. 7. VWS main view with part description: 1 - one of the clips (is fastened on the base butt-end), 2 - the sensor base made of material with low coefficient of thermal expansion, 3, 5 – magnet poles system, 4 – dielectric plates, 6 - vibrating wire, 7, 8 – fasteners.

Fig. 8. Laser beam scanning by vibrating wire.

Fig. 9. Frequency-temperature dependence of wires before (a) and after (b) irradiation.

Fig. 10. Horizontal profile of an electron beam with a current of 10 nA scanned with a beryl-bronze wire of 90 µm diameter as the vibrating wire sensor.

Fig. 11. VWS under ion beam of EMAL-2.

Fig. 12. VWS in park position in PETRA vacuum chamber. Wire distance from beam center is $6.7\sigma$ (40 mm).

Fig. 13. The VWS output signal in park position.

Fig. 14. Beam scanning at full feed with different vertical shifts of the beam: 1 – the frequency signal from VWS, 2 – position of the VWS, 3 – beam current, 4 – beam losses.

Fig. 15. Scanning of the beam with fixed bumps. Curves: 1 – frequency signal from VWS, 2 – VWS position, 3 – beam current, 4 - beam losses.

Fig. 16. Noises of frequency signal from VWS in stationary position.

Fig. 17. Some shallow scans. Curves: 1 – frequency signal from VWS, 2 – VWS position, 3 – beam current, 4 – PM1 countings.

Fig. 18. Scan of the proton beam using VWS: 1- frequency of the VWS, 2 – beam current, 3 – VWS position relative to the vacuum chamber center, 4 and 5 – signals from scintillators.

Fig. 19. Comparison of signals from VWS with PM1(curve 1) and PM2 (curve 2).

**Figure 1**

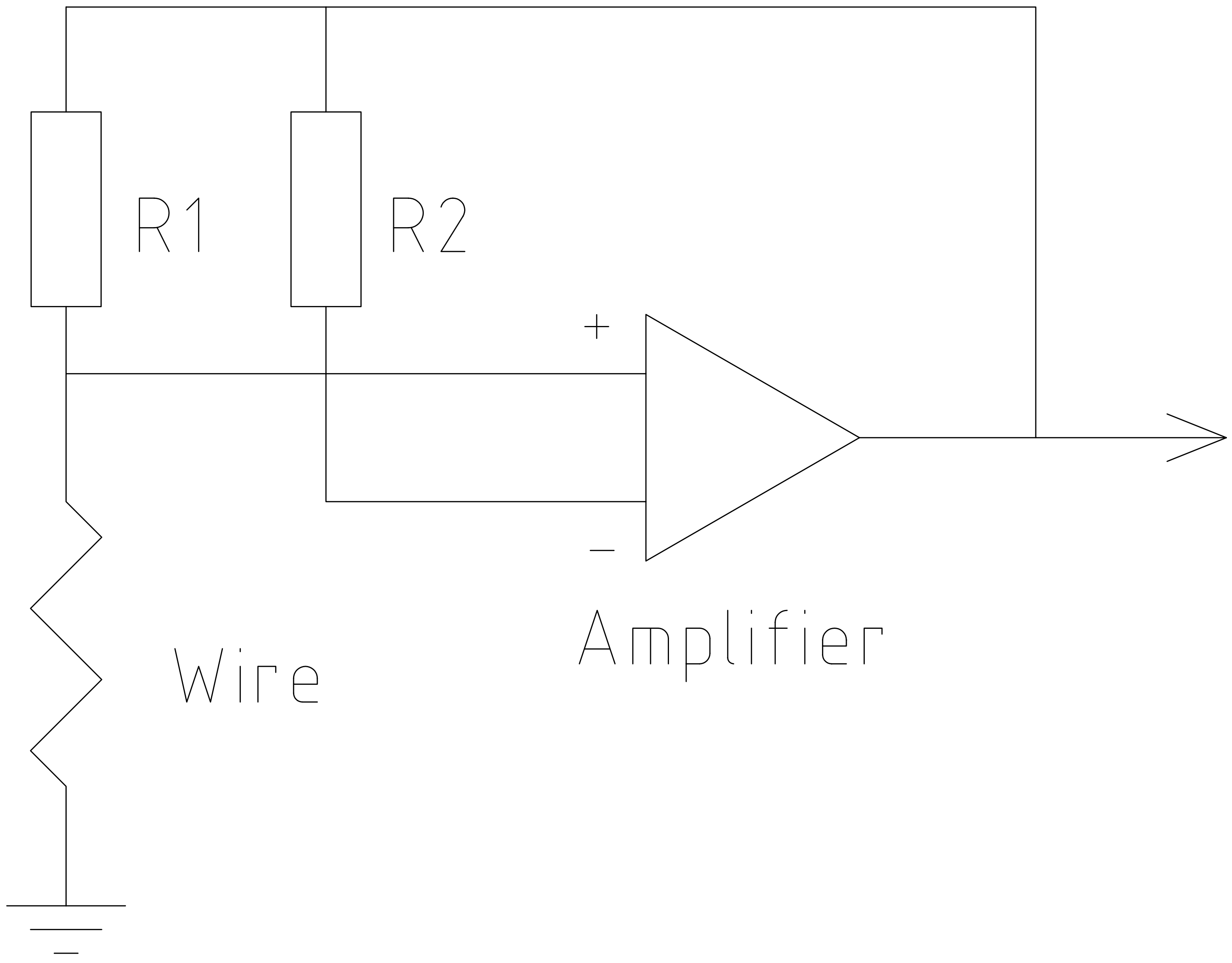

Figure 2

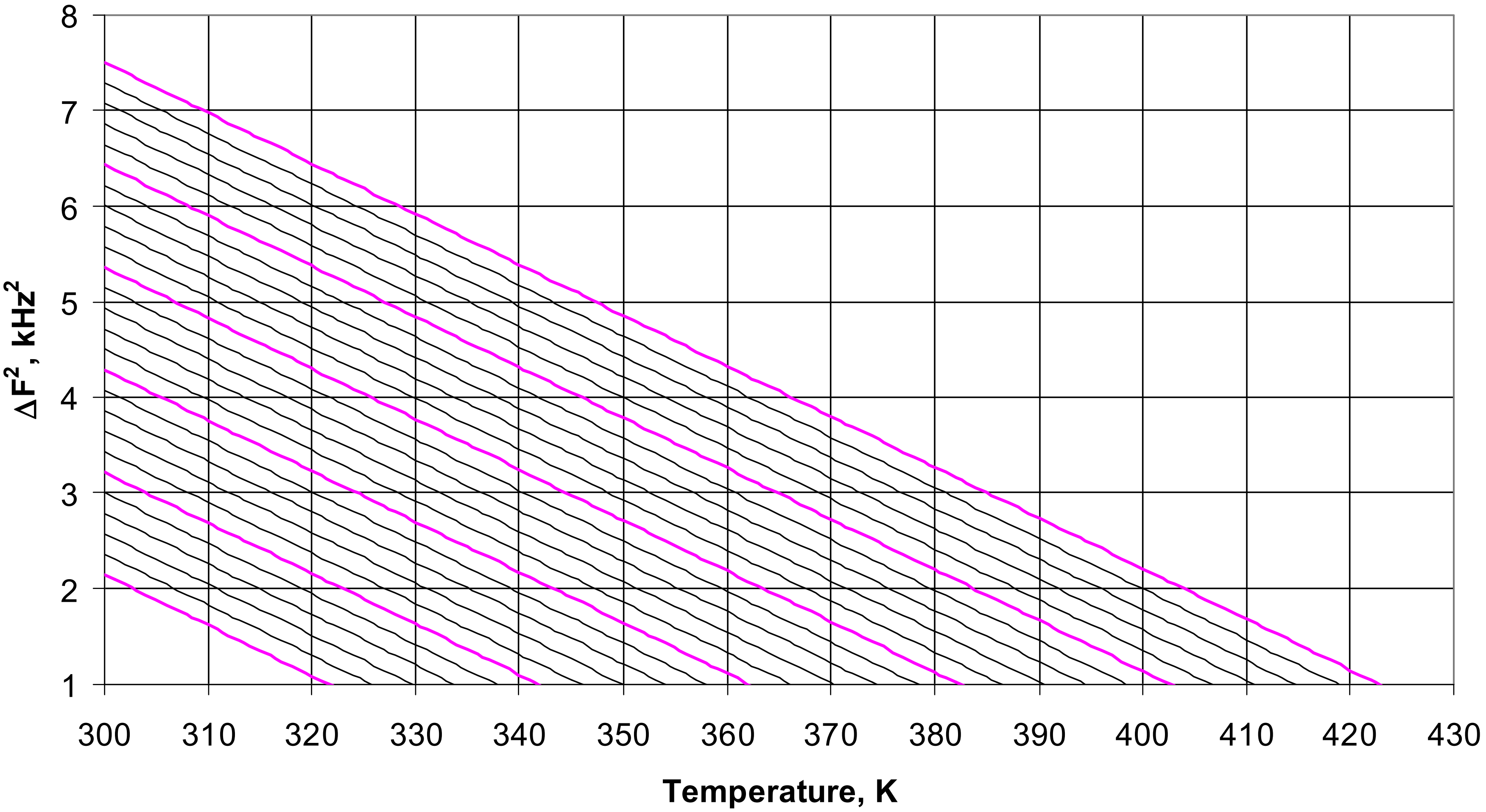

Figure 3

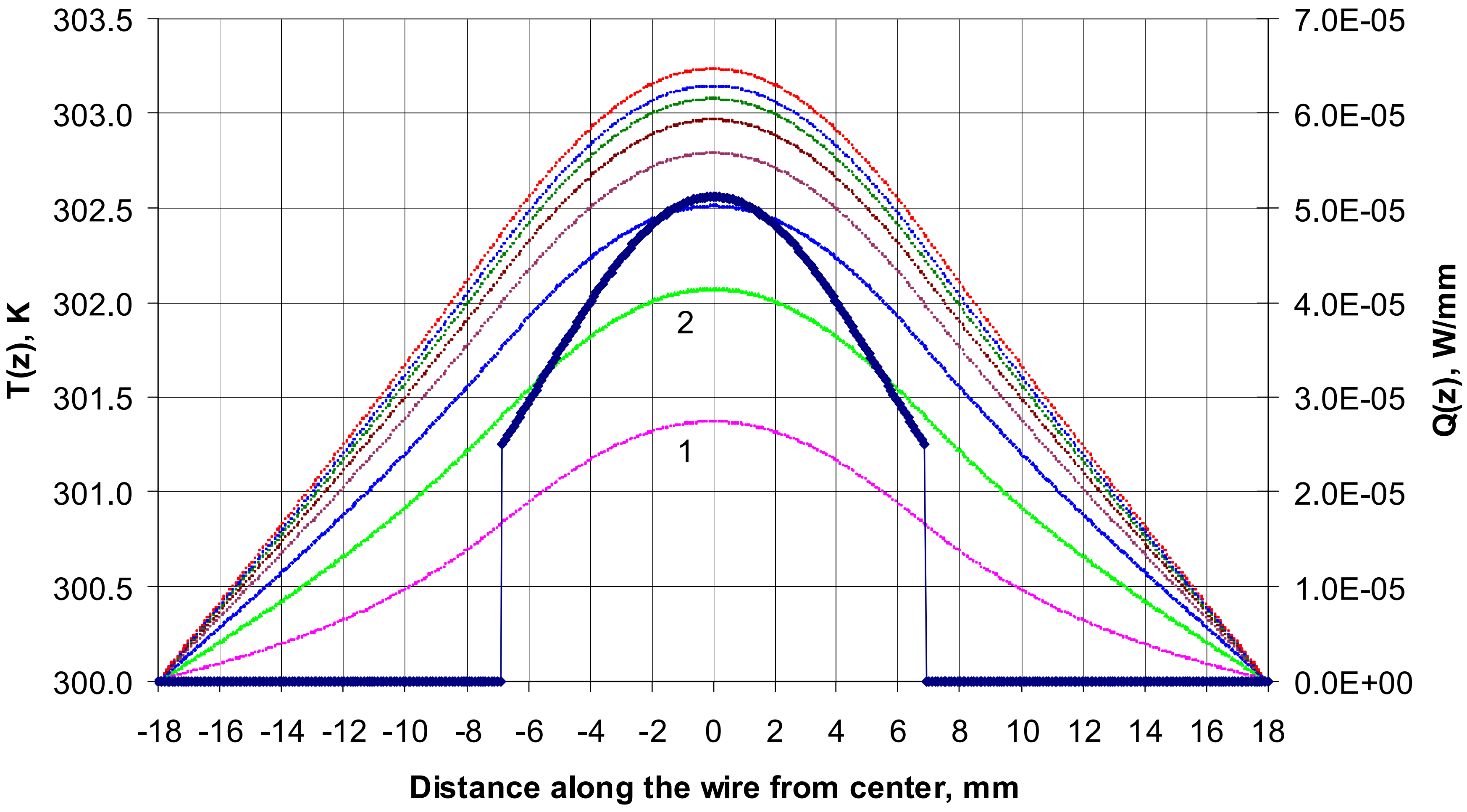
303.5
303.0
302.5
302.0
301.5
301.0
300.5
300.0
T(z), K
7.0E-05
6.0E-05
5.0E-05
4.0E-05
3.0E-05
2.0E-05
1.0E-05
0.0E+00
Q(z), W/mm
2
1
-18 -16 -14 -12 -10 -8 -6 -4 -2 0 2 4 6 8 10 12 14 16 18
Distance along the wire from center, mm

Figure 4

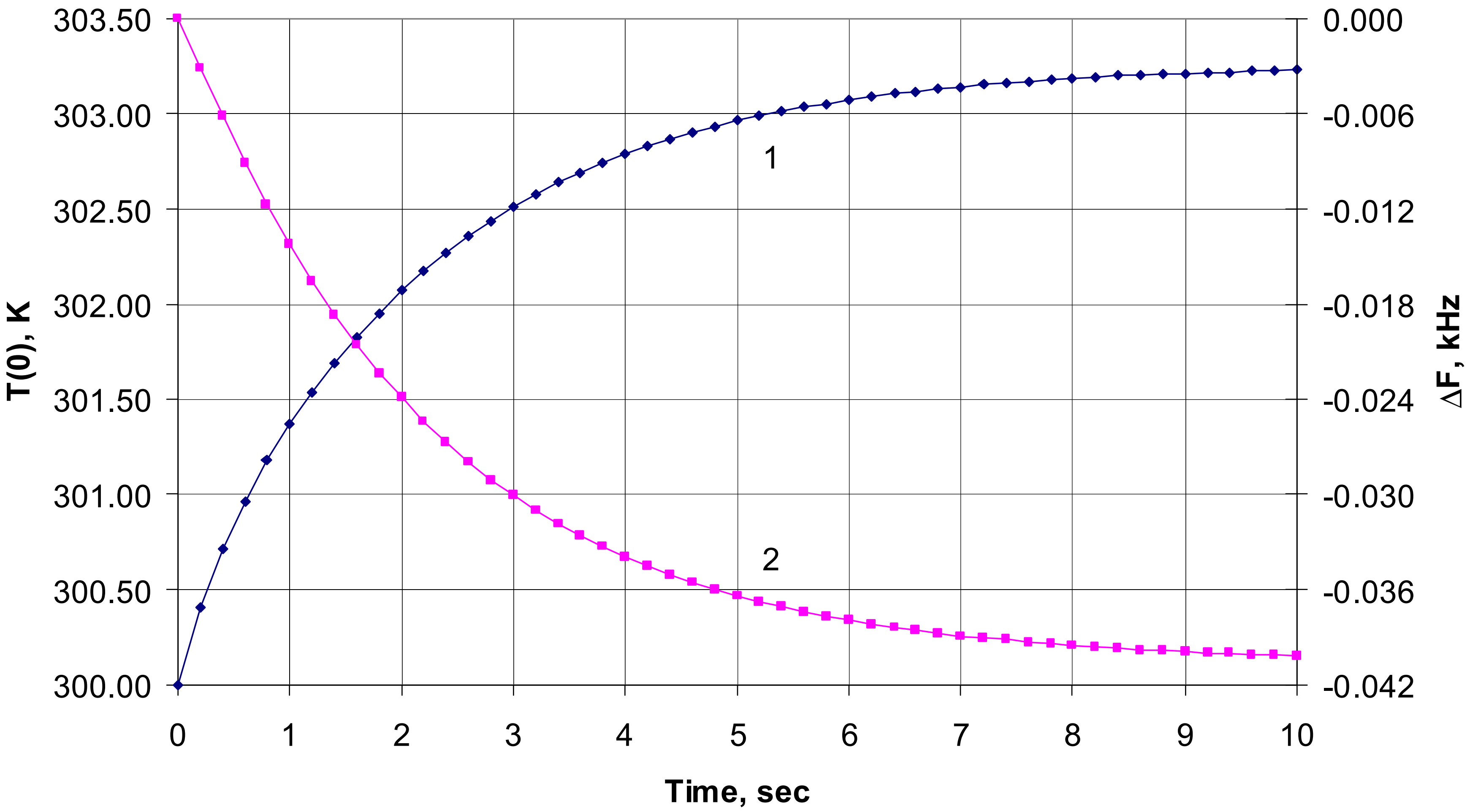

Figure 5

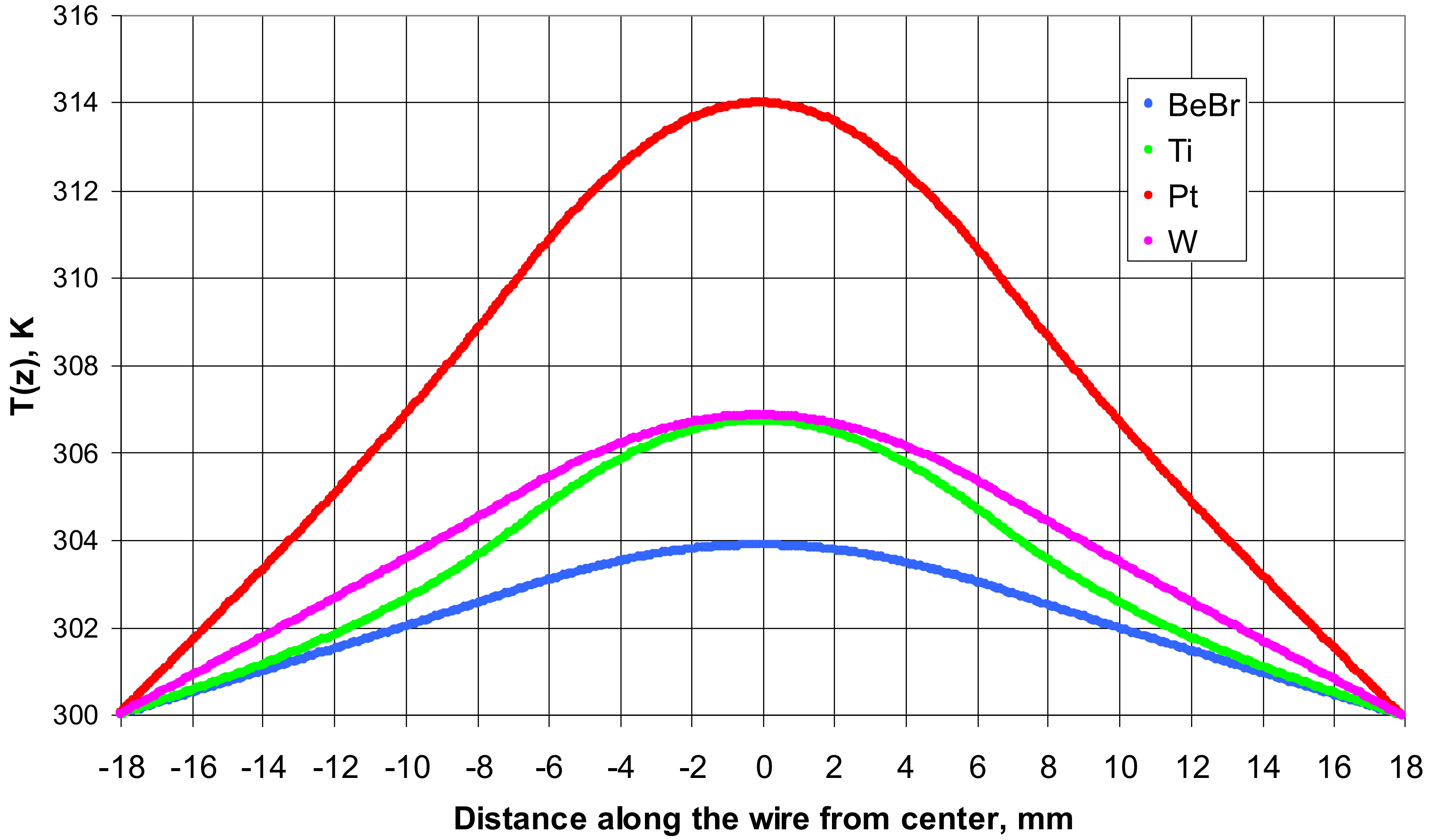

Figure 6a

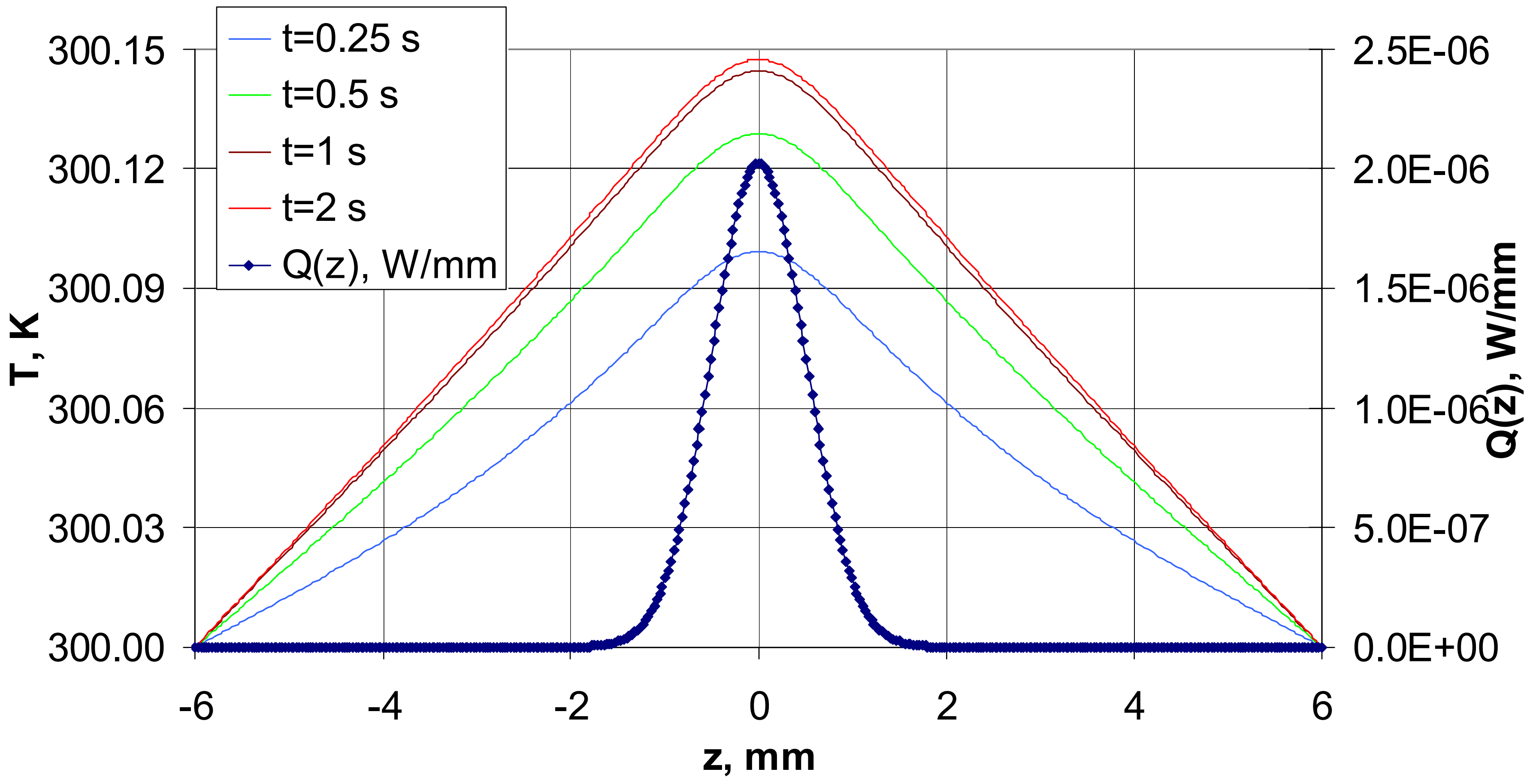

Figure 6b

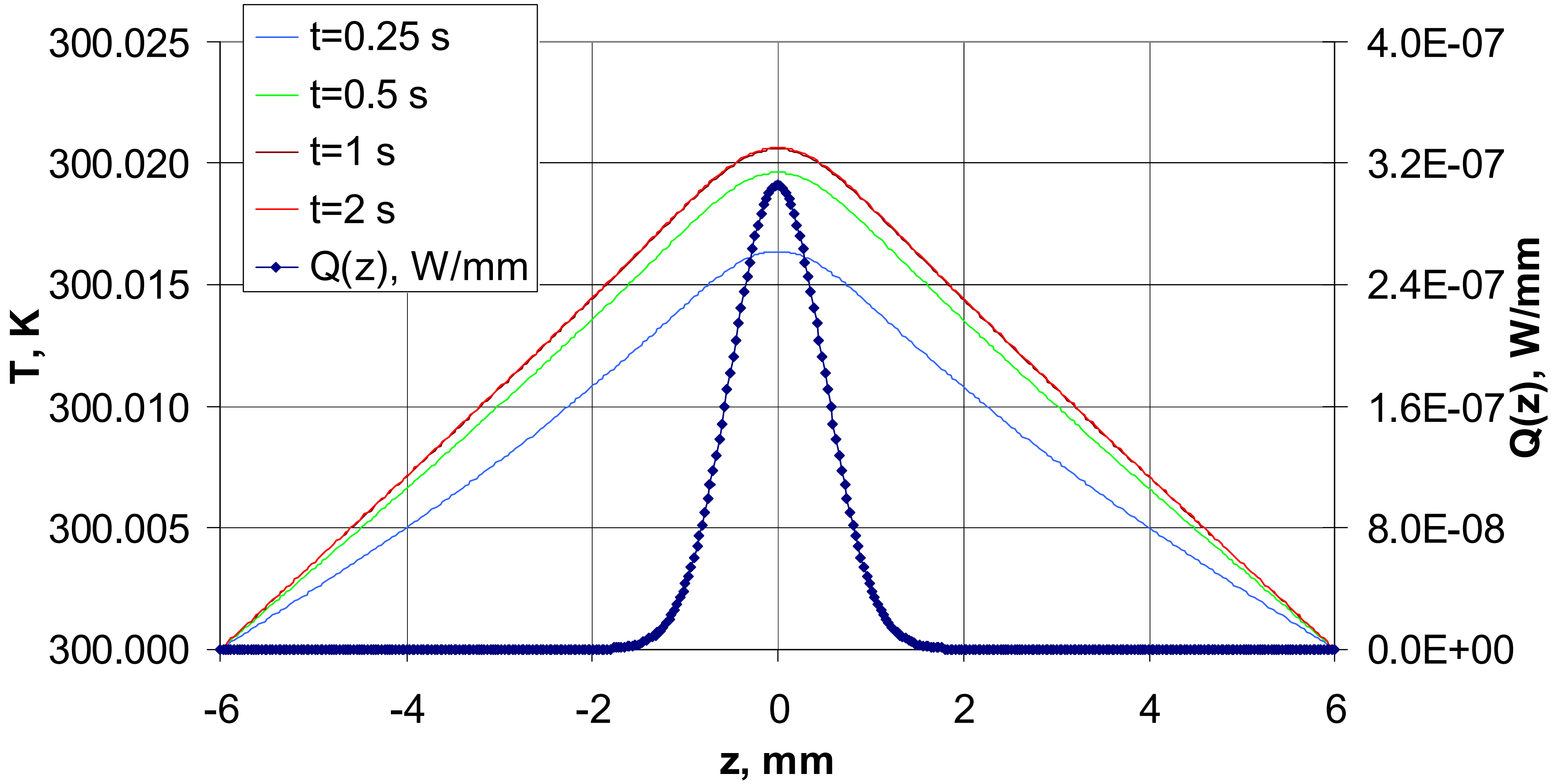

Figure 7
Click here to download high resolution image

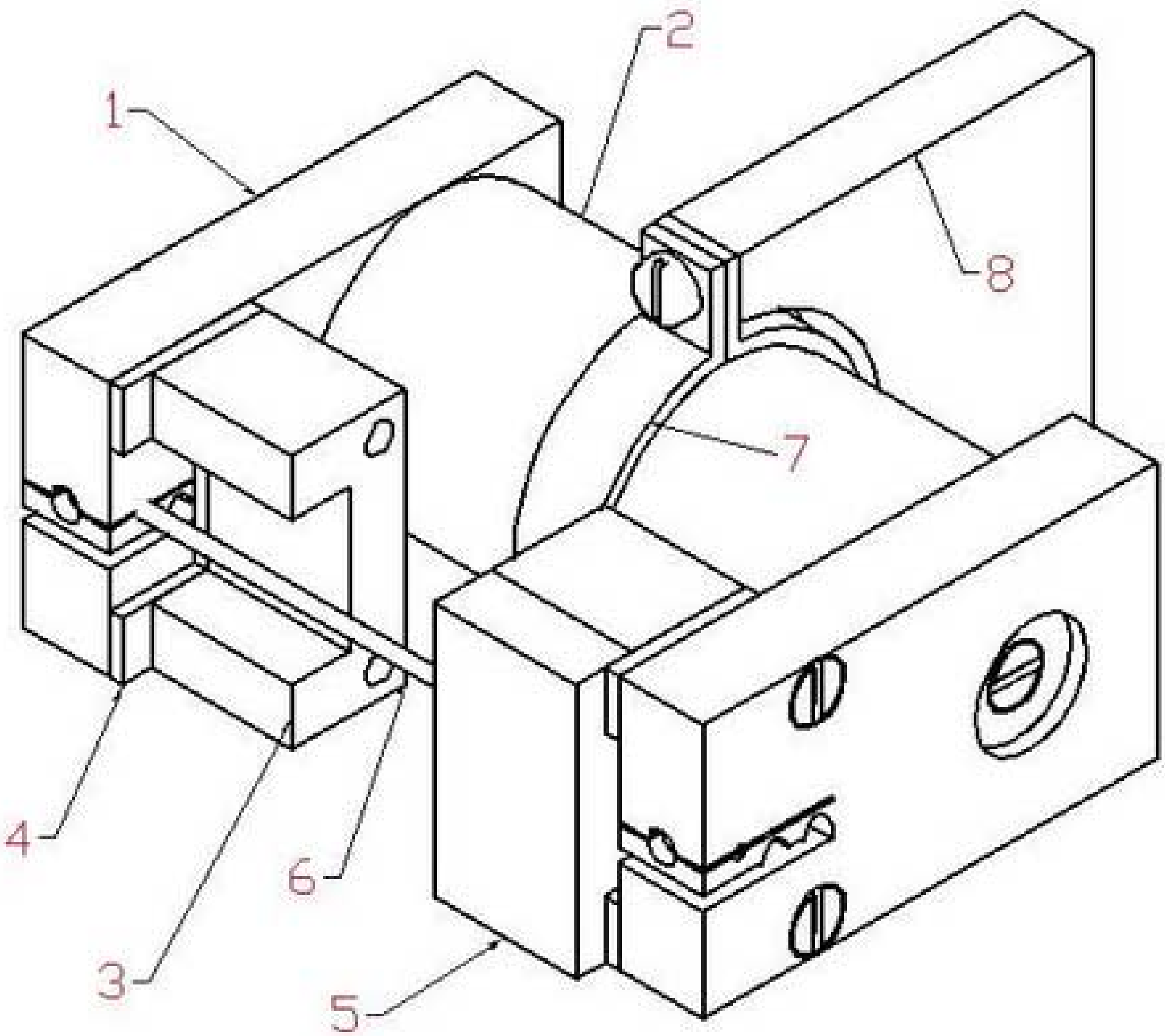

Figure 8

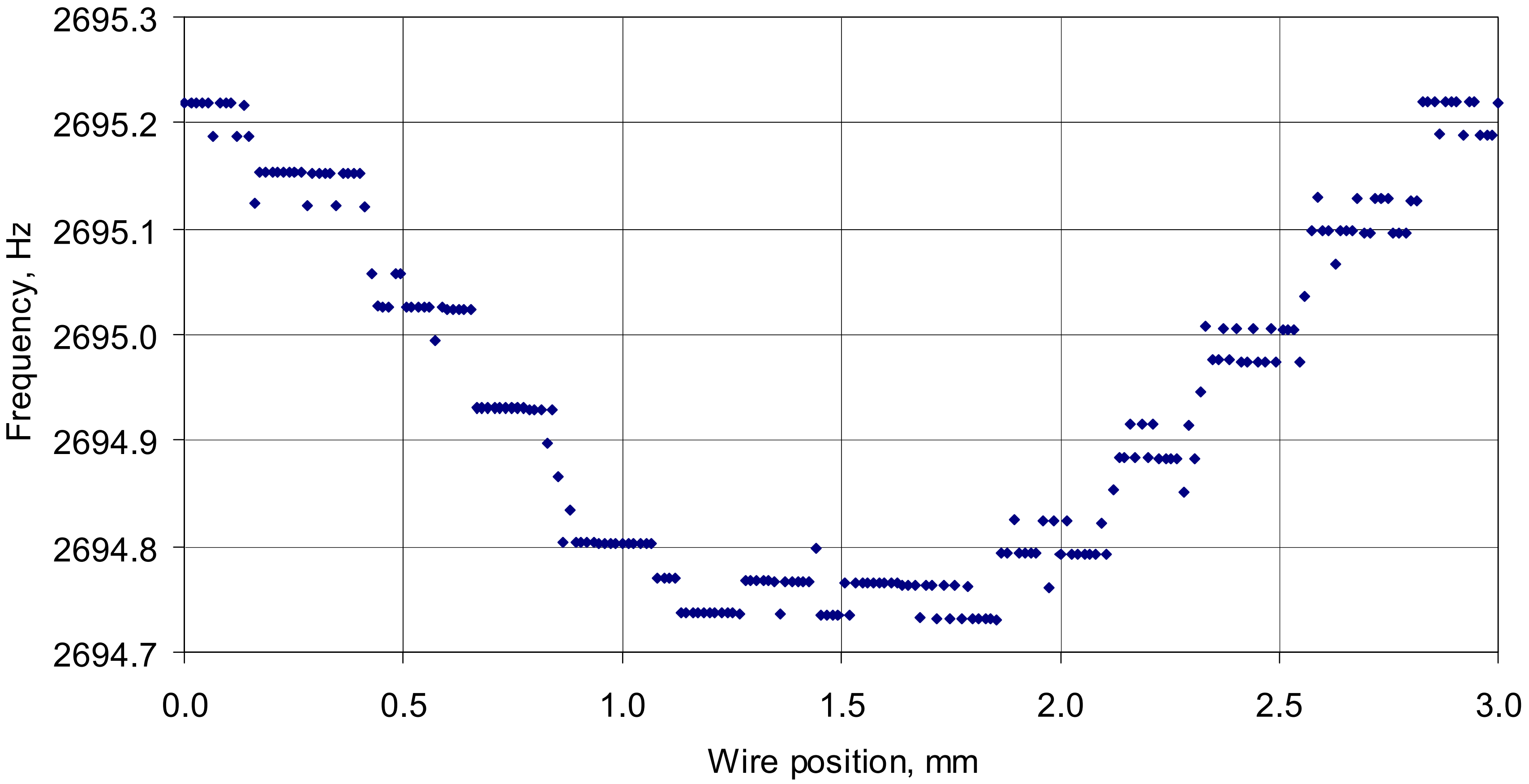

Figure 9a

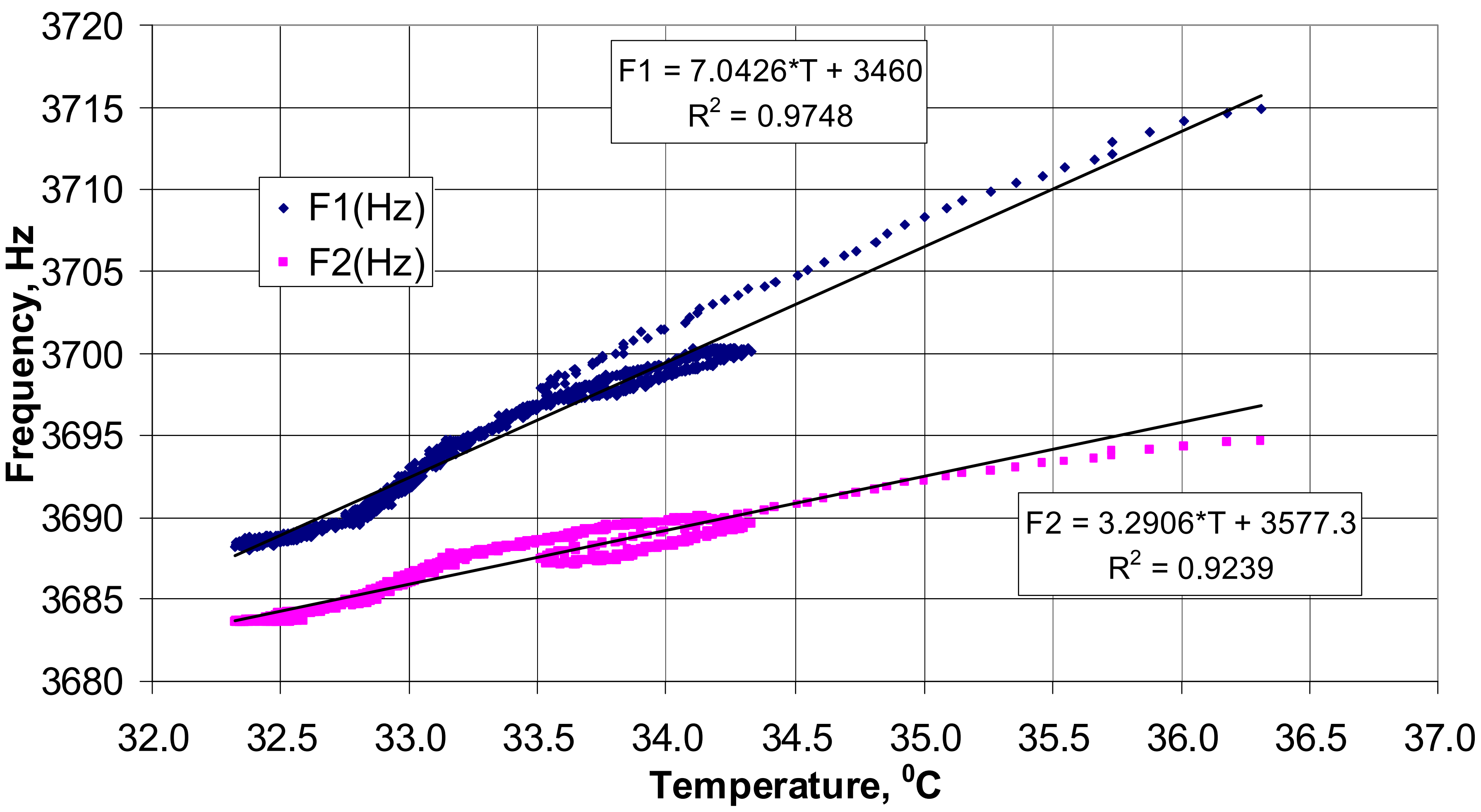

Figure 9b

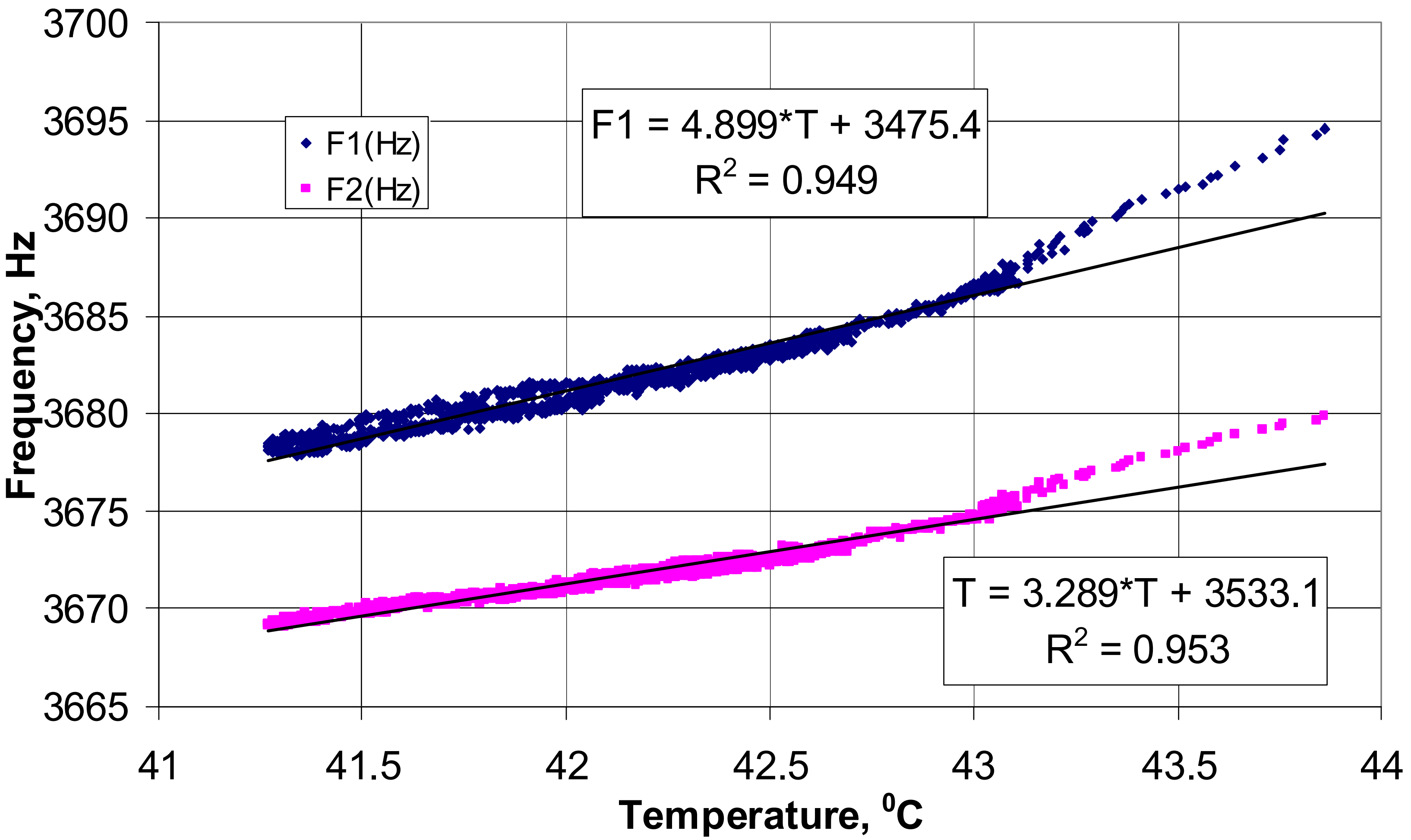

Figure 10

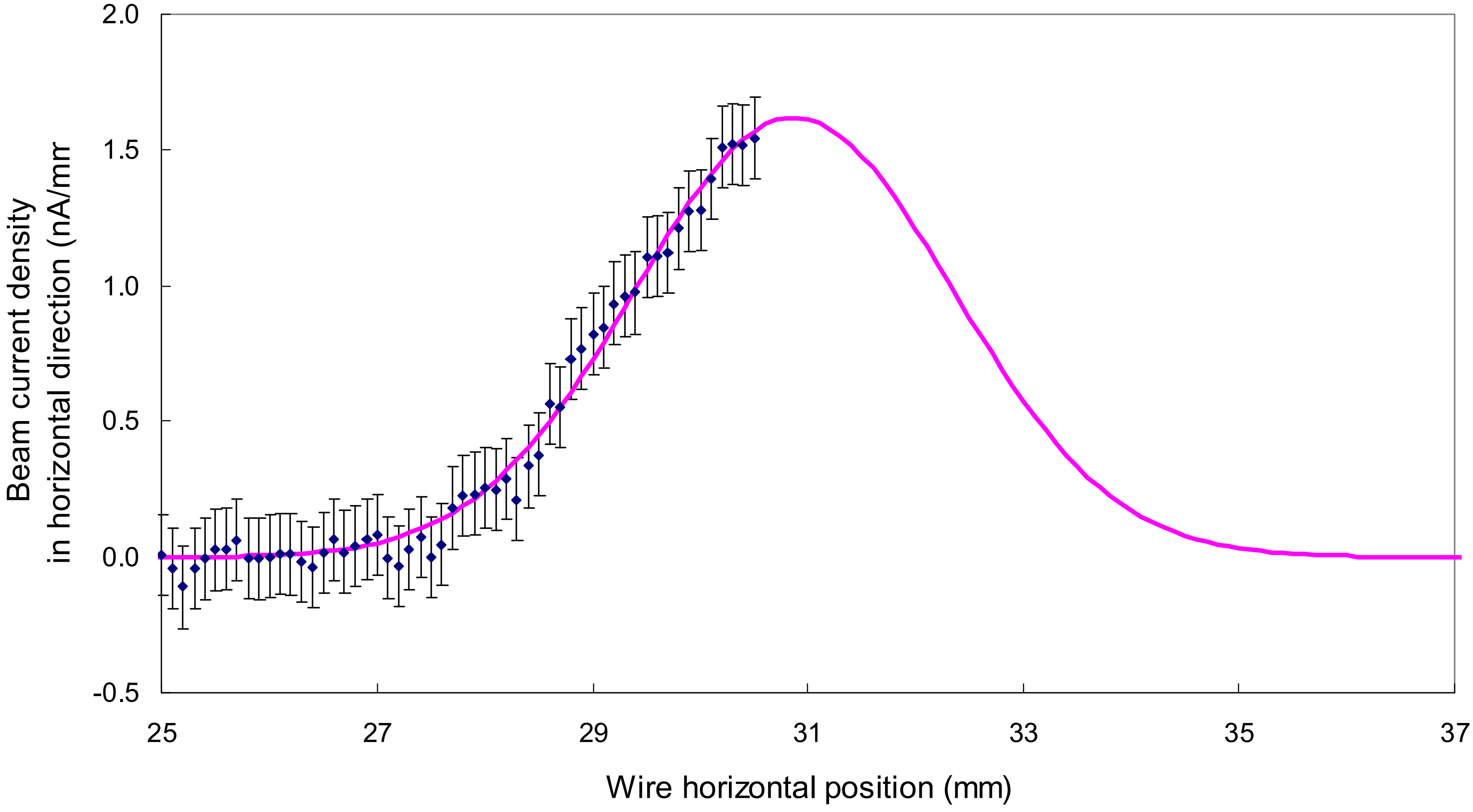

Figure 11

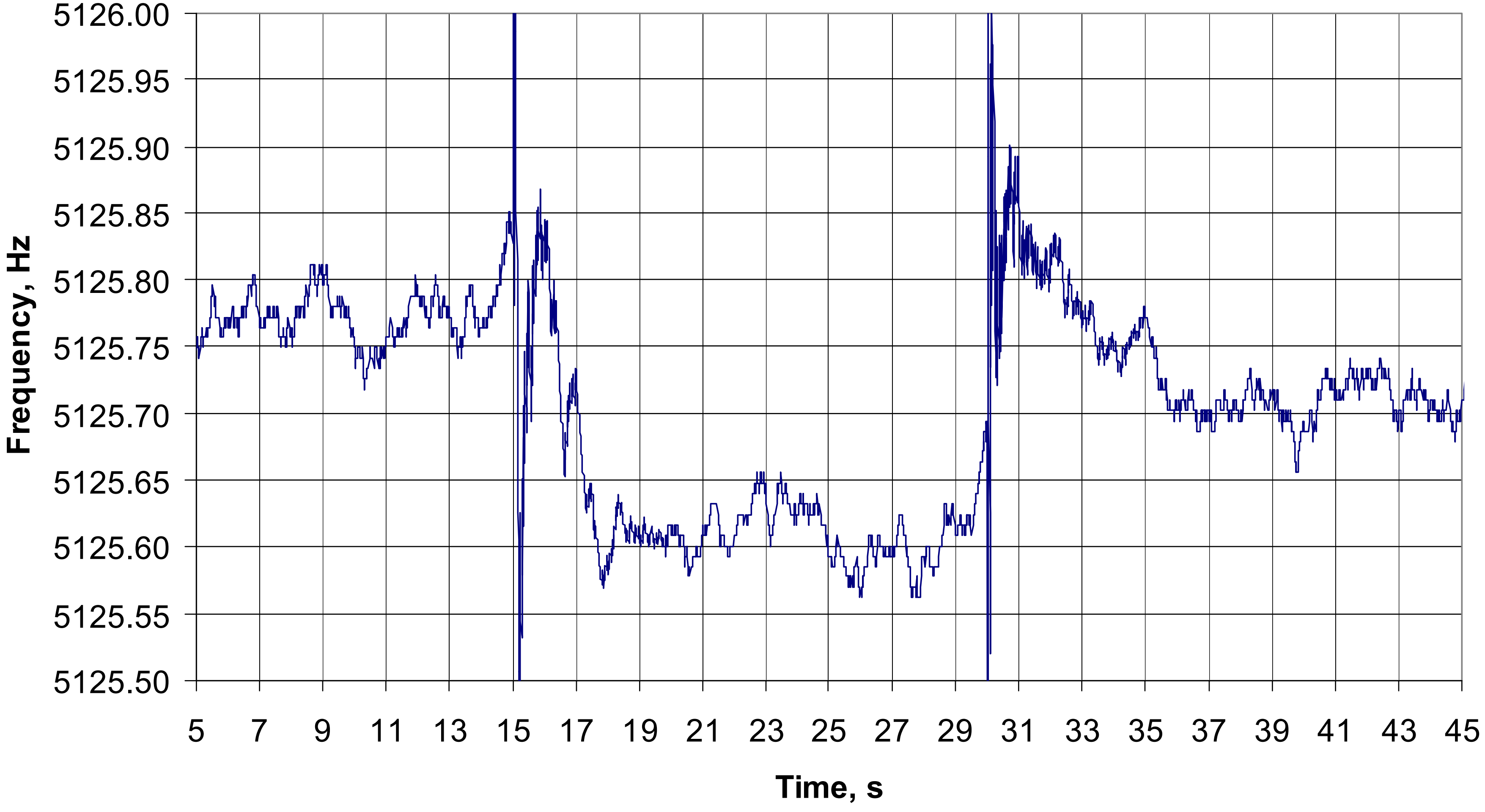

Figure 12
Click here to download high resolution image

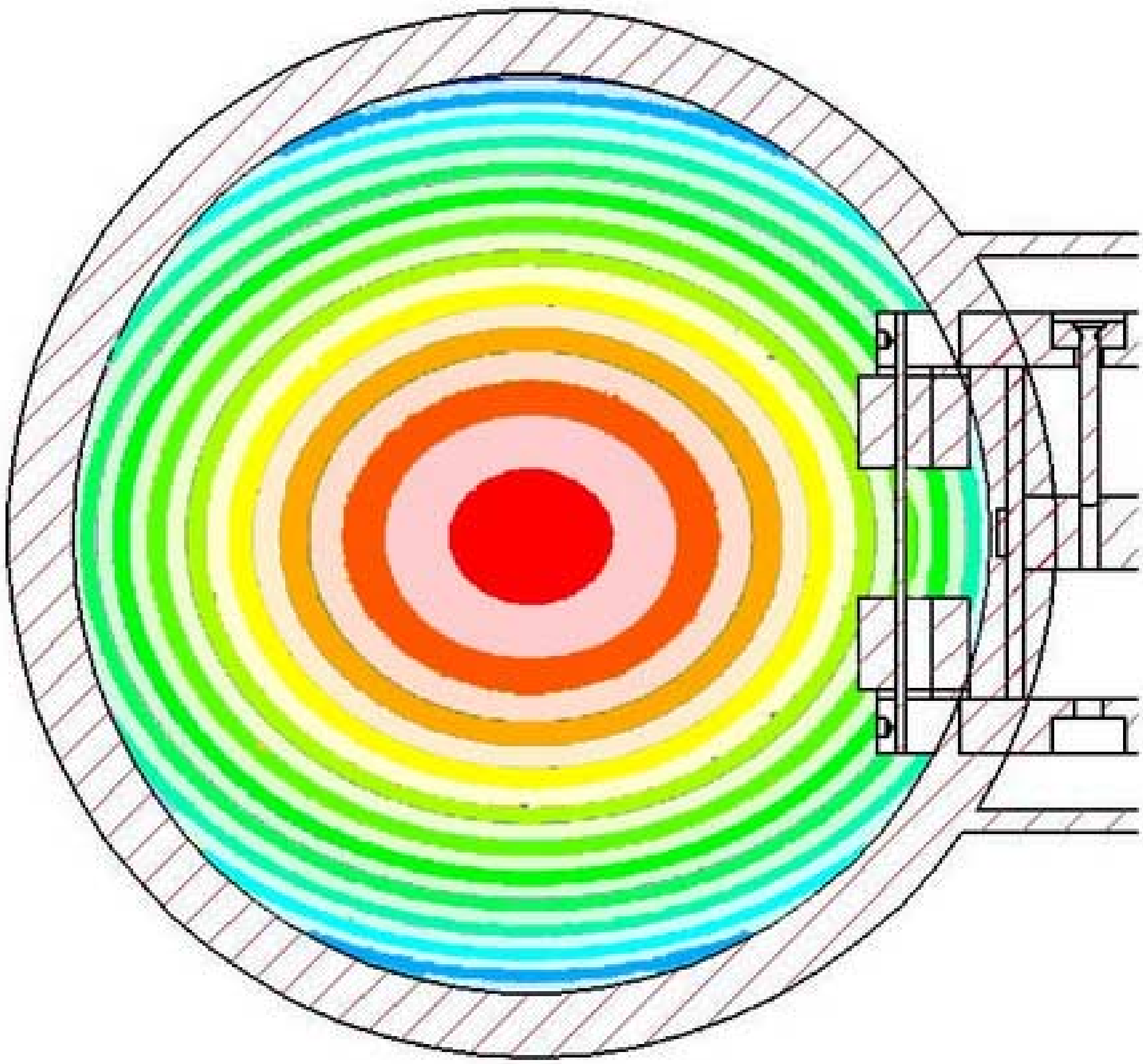

Figure 13

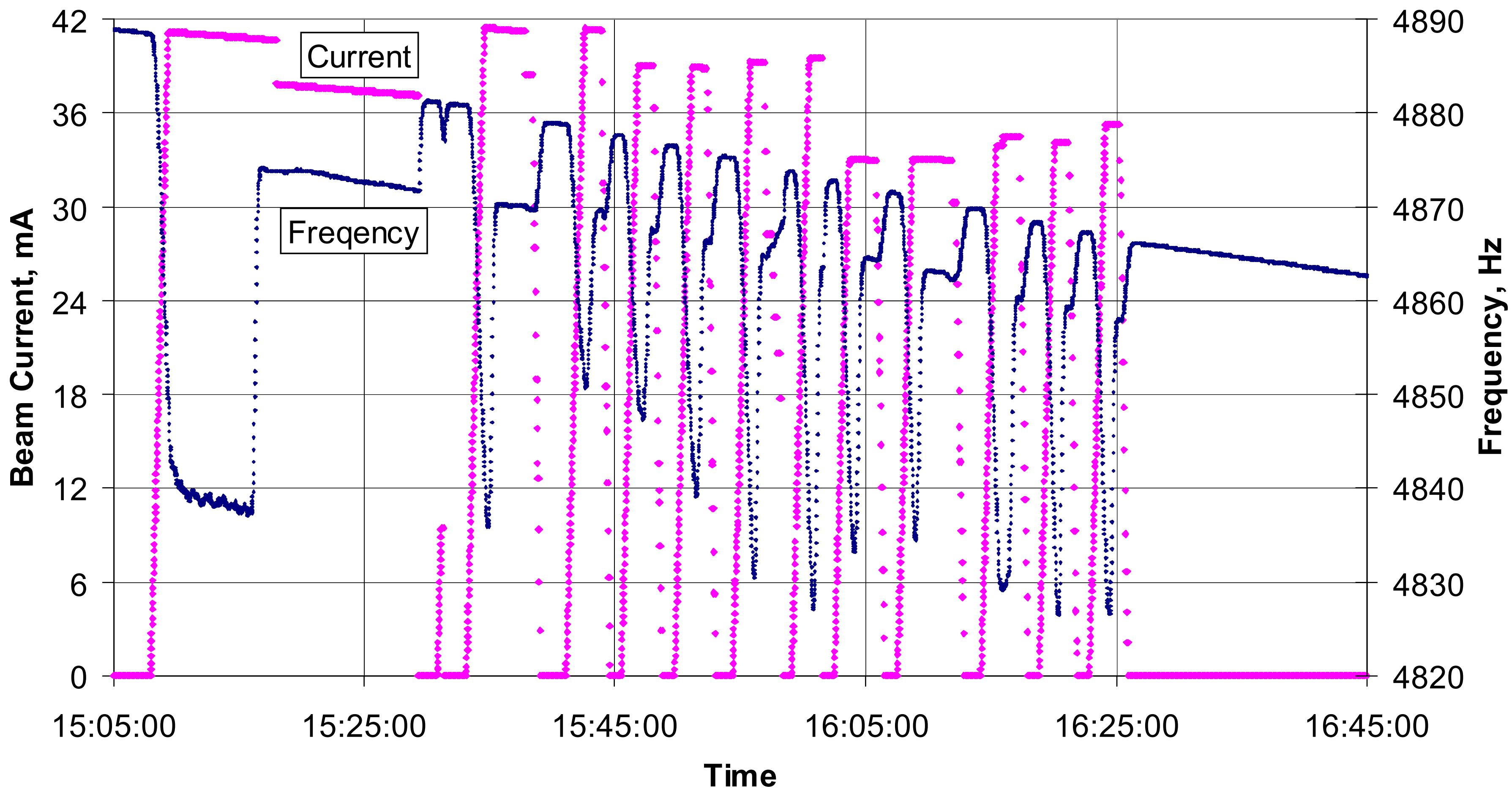

Figure 14

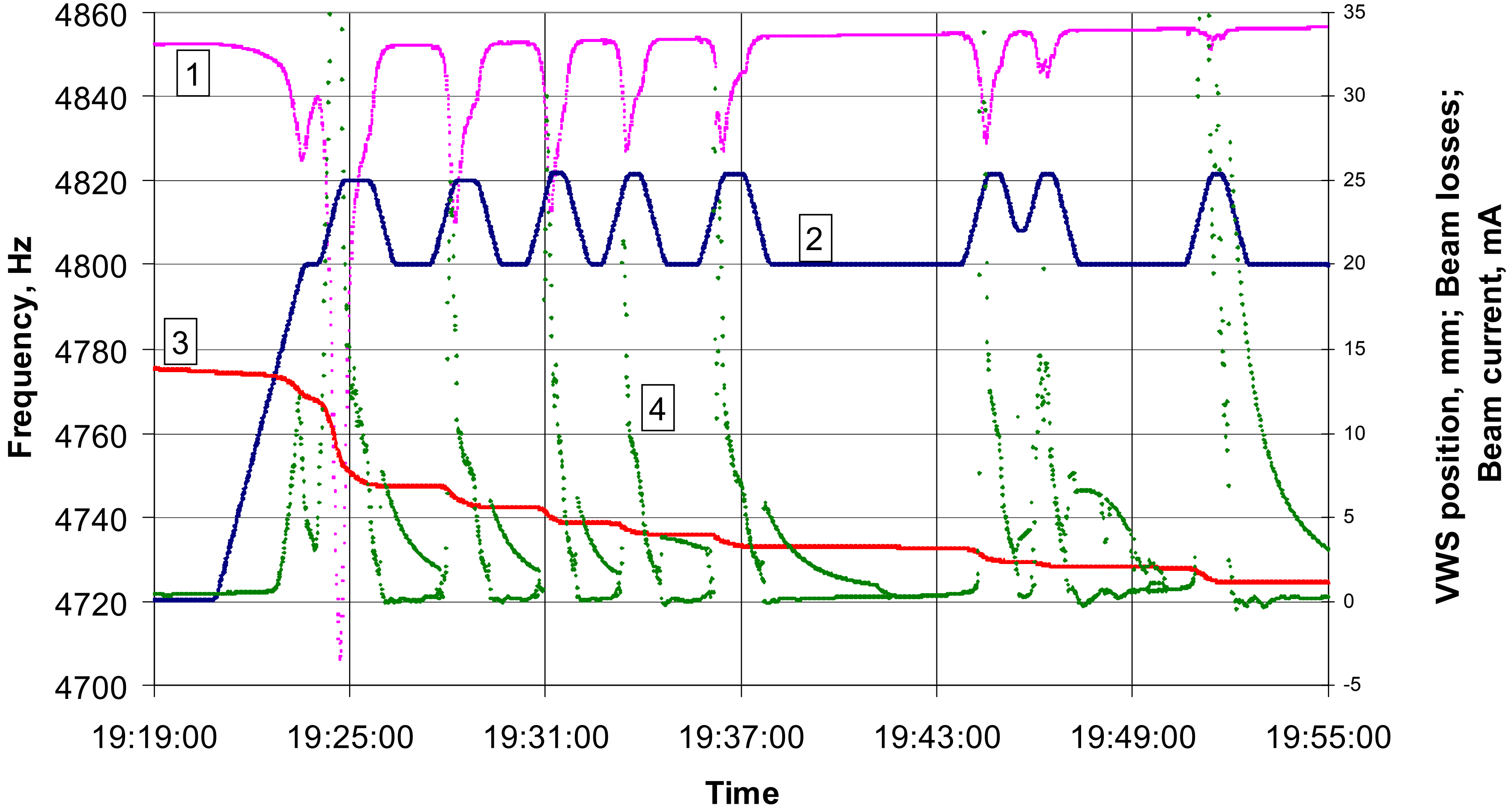

4860
4840
4820
4800
4780
4760
4740
4720
4700
Frequency, Hz
35
30
25
20
15
10
5
0
-5
VWS position, mm; Beam losses; Beam current, mA
1
2
3
4
19:19:00
19:25:00
19:31:00
19:37:00
19:43:00
19:49:00
19:55:00
Time

Figure 15

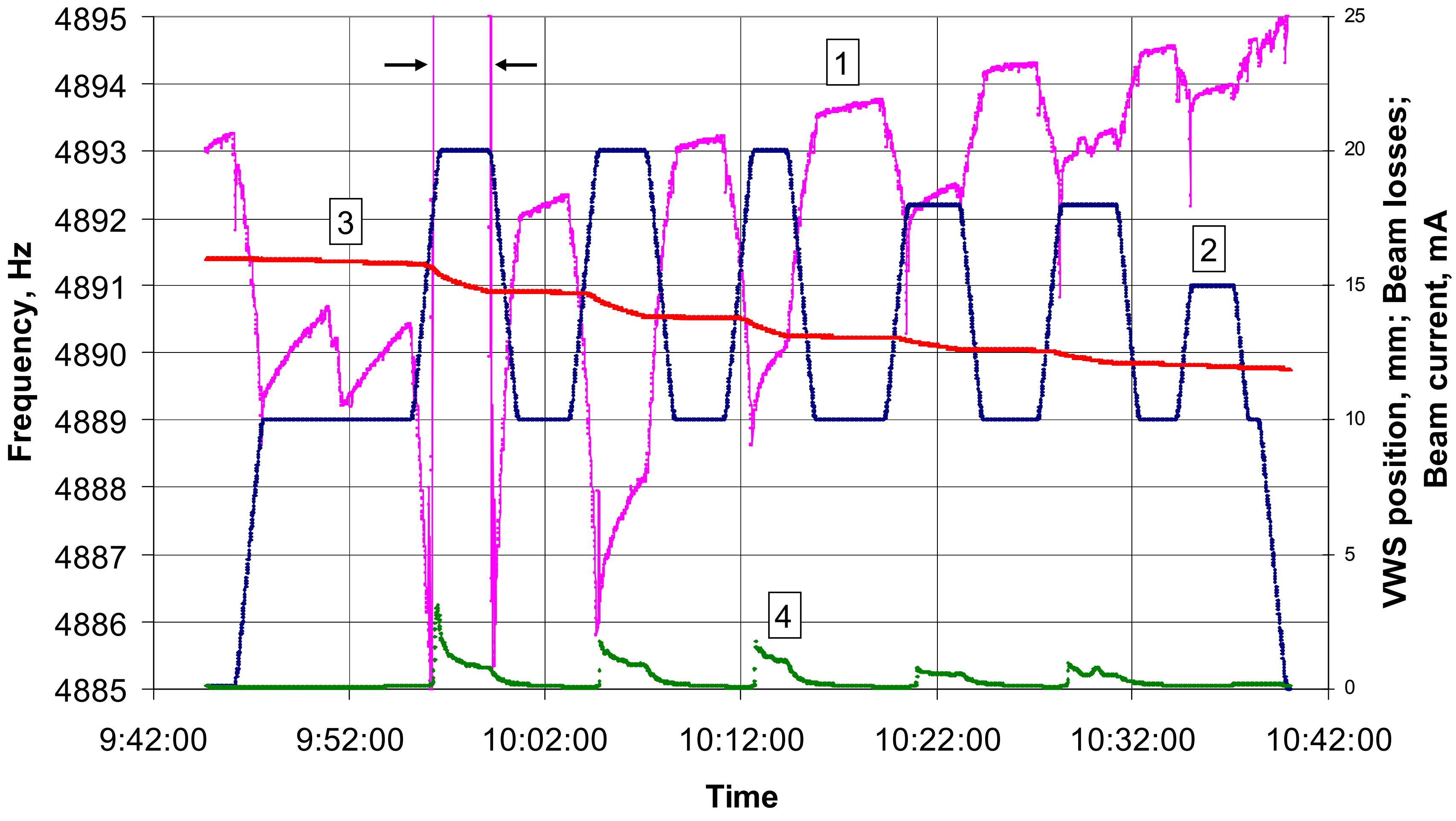

Figure 16

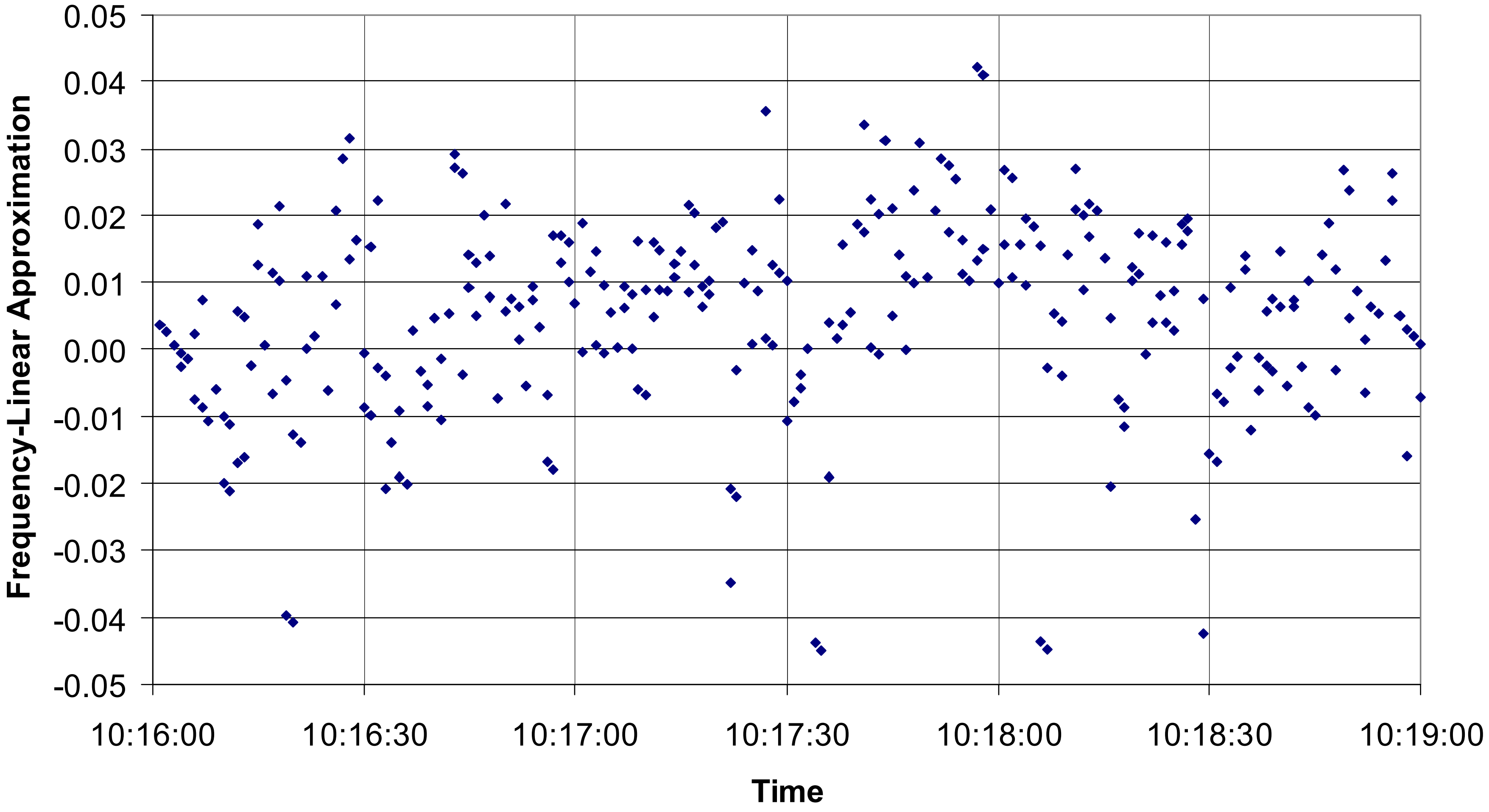

Figure 17

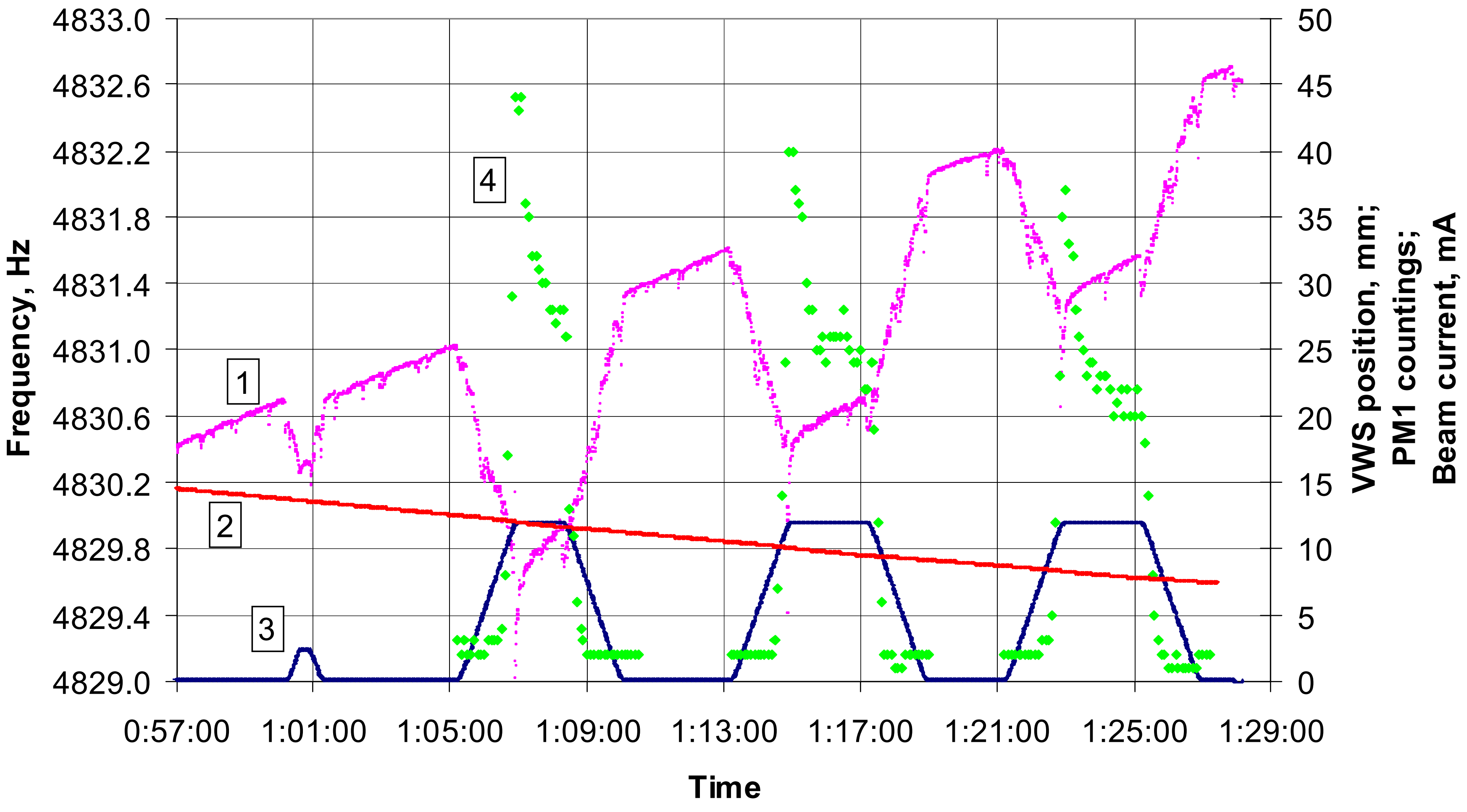

4833.0
4832.6
4832.2
4831.8
4831.4
4831.0
4830.6
4830.2
4829.8
4829.4
4829.0
Frequency, Hz
50
45
40
35
30
25
20
15
10
5
0
VWS position, mm;
PM1 countings;
Beam current, mA
0:57:00
1:01:00
1:05:00
1:09:00
1:13:00
1:17:00
1:21:00
1:25:00
1:29:00
Time
1
2
3
4

Figure 18

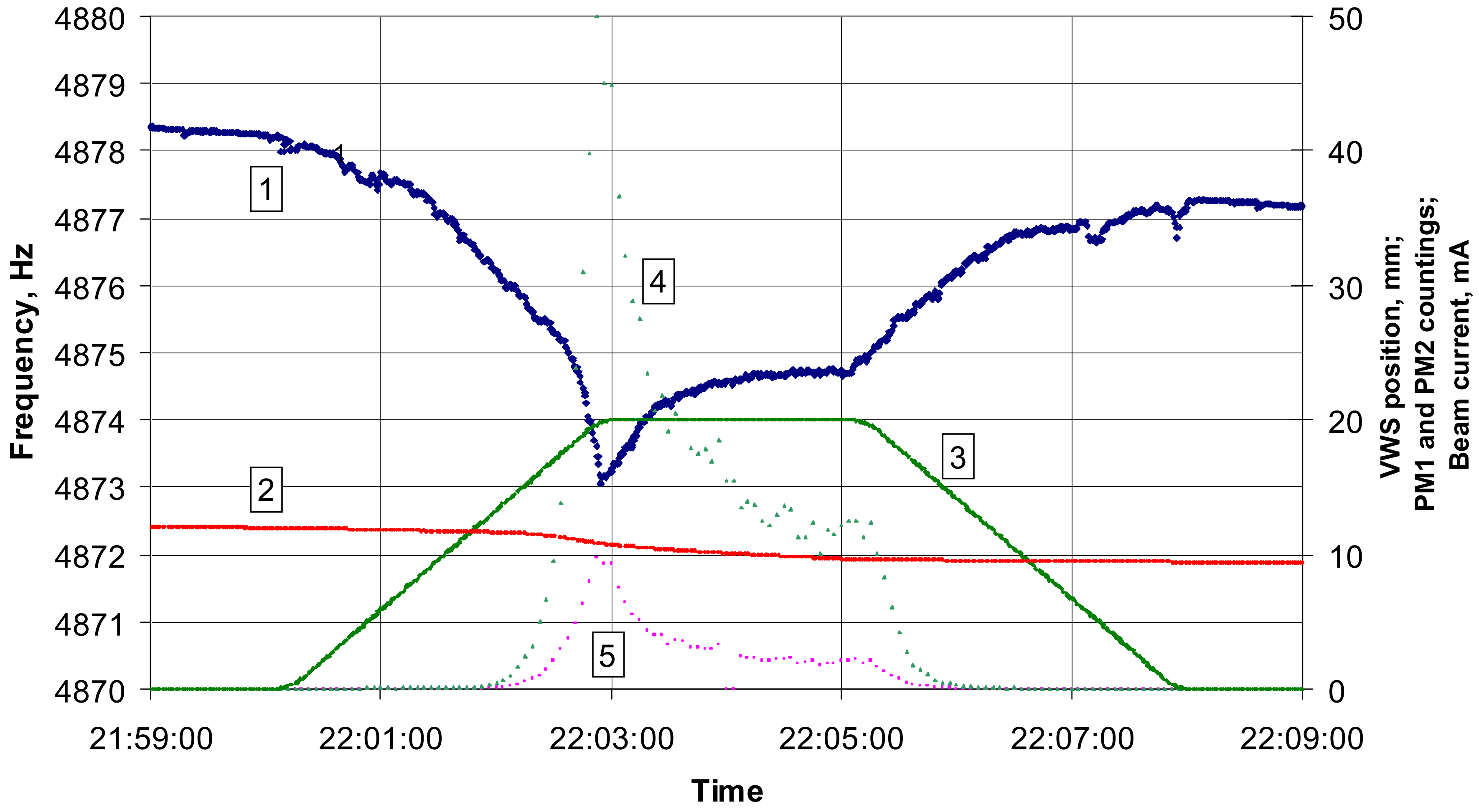

Frequency, Hz
4880
4879
4878
4877
4876
4875
4874
4873
4872
4871
4870
50
40
30
20
10
0
VWS position, mm;
PM1 and PM2 countings;
Beam current, mA
1
2
3
4
5
21:59:00
22:01:00
22:03:00
22:05:00
22:07:00
22:09:00
Time

Figure 19

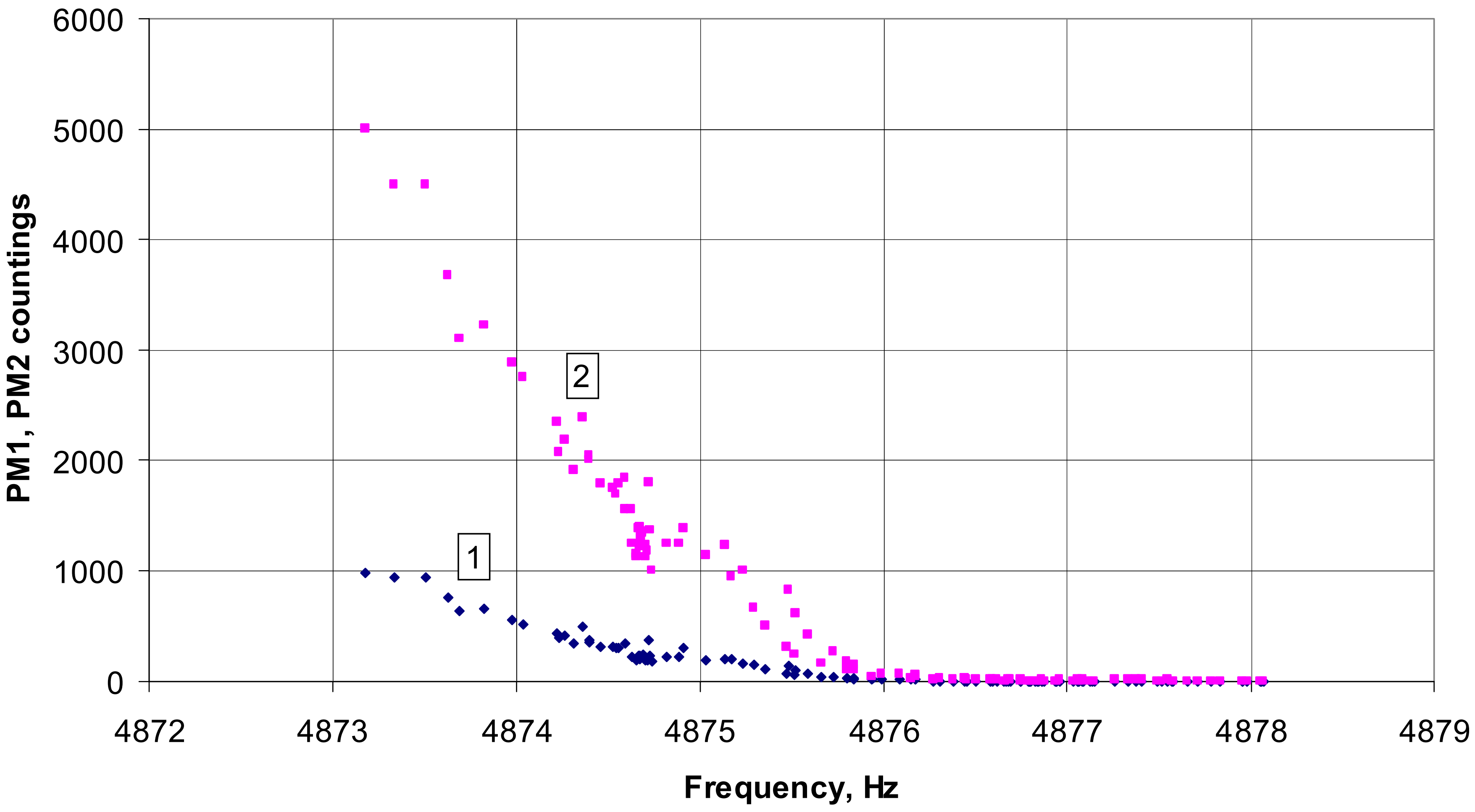

**Table 1**

Table 1. Temperature range of vibrating wire sensor

| *Material* | *E, GPa* | $\alpha_s$, *1/K* | $\sigma_{0.2}$, *GPa* | $\Delta T_S$, *K* |
|---|---|---|---|---|
| Beryl Bronze hard | 130 | 1.90E-5 | 0.9* | 482 |
| Wolfram Recrystal | 400 | 4.70E-6 | 0.5* | 433 |
| Titan ($\sigma_{0.2}$ at $20^0$ C) | 110 | 9.86E-6 | 0.66 | 909 |
| Titan, therm. treated, pure | 110 | 9.86E-6 | 0.1 | 392 |
| Titan, therm. treated, 99.6% | 110 | 9.86E-6 | 0.3 | 577 |
| Platinum, therm. treated | 160 | 9.70E-6 | 0.07 | 341 |
| SiC, fiber | 400 | 4.50E-6 | 2* | 856 |
| SiC, crystal (whiskers) | 580 | 4.60E-6 | 21* | 4240 |
| SiC, (whiskers) | 450 | 4.60E-6 | 3.05* | 1770 |
| $SiO_2$, fiber | 550 | 6.25E-6 | 1.38* | 702 |

**Table 2**

Table 2. Some typical parameters of wire heating process by proton beam

| x, mm | $I_s(x)$, A | $I_s(x)/I_s(0)$ | $Q_s$, W | $T_{mean}$, K | Δf, Hz |
|---|---|---|---|---|---|
| 13 | 5.066E-06 | 1.019E-01 | 1.395E-01 | 3.45E+02 | |
| 17 | 1.001E-06 | 2.014E-02 | 2.760E-02 | 6.82E+01 | 1.473E+03 |
| 20 | 2.238E-07 | 4.494E-03 | 6.160E-03 | 1.52E+01 | 3.287E+02 |
| 25 | 1.069E-08 | 2.149E-04 | 2.945E-04 | 7.28E-01 | 1.572E+01 |
| 30 | 2.592E-10 | 5.229E-06 | 7.169E-06 | 1.77E-02 | 3.825E-01 |
| 35 | 3.214E-12 | 6.475E-08 | 8.869E-08 | 2.19E-04 | 4.736E-03 |
| 40 | 2.020E-14 | 4.079E-10 | 5.587E-10 | 1.38E-06 | 2.984E-05 |

**Table 3**

Table 3. Charged particles energy losses through the radiation mechanism and thermal conductivity

| Material | $Q_{rad}$, W | $Q_{\lambda}$, W | T(0), K | ΔF, kHz |
|---|---|---|---|---|
| Beryl Bronze | 1.156E-04 | 4.427E-04 | 3.032E+02 | -4.02E-02 |
| Titan | 1.843E-04 | 6.554E-05 | 3.058E+02 | -5.28E-02 |
| Platinum | 4.270E-04 | 6.300E-04 | 3.119E+02 | -4.04E-02 |
| Tungsen | 2.105E-04 | 8.170E-04 | 3.058E+02 | -2.56E-02 |

**Table 4**

Table 4. Absorption parameter $\mu^{-1}$ ($cm^{-1}$) for photon beams with different energies.

| Photon energy, eV | Titan | Silicium | Molybdenum | Tungsten |
|---|---|---|---|---|
| 1.0E+02 | 1.530E+05 | 7.445E+04 | 5.038E+04 | 3.541E+05 |
| 5.0E+02 | 1.116E+05 | 2.323E+04 | 1.951E+05 | 2.241E+05 |
| 1.0E+03 | 2.629E+04 | 3.733E+03 | 5.324E+04 | 7.629E+04 |
| 5.0E+03 | 3.191E+03 | 5.550E+02 | 5.622E+03 | 1.070E+04 |
| 1.0E+04 | 5.088E+02 | 7.454E+01 | 8.485E+02 | 1.800E+03 |
| 5.0E+04 | 1.018E+02 | 1.491E+01 | 1.697E+02 | 3.599E+02 |